\documentclass[11pt]{article}

\usepackage[preprint]{acl}

\usepackage{times}
\usepackage{latexsym}

\usepackage[T1]{fontenc}

\usepackage[utf8]{inputenc}

\usepackage{microtype}

\usepackage{inconsolata}

\usepackage{graphicx}

\usepackage{amsmath}

\usepackage{makecell}

\usepackage{booktabs}

\usepackage[table]{xcolor}

\usepackage{multirow}

\usepackage{amssymb}

\usepackage[most]{tcolorbox}
\tcbuselibrary{breakable}

\title{CoDeR: Local Constraint-Compatible Retrieval Beyond Semantic Similarity}

\author{
  Xingkun Yin$^1$ \quad Xuebin Tang$^2$ \quad \textbf{Hongyang Du}$^1$\thanks{Corresponding author}\\
  $^1$Department of Electrical and Computer Engineering,
  University of Hong Kong \\
  Hong Kong SAR, China \\
  $^2$School of Computer and Communication Engineering, 
  University of Science and Technology Beijing \\
  Beijing, China \\
  \{yinxingkun@connect., duhy@eee.\}hku.hk \quad tangxb@xs.ustb.edu.cn\\
}

\begin{document}
\maketitle
\begin{abstract}
Information retrieval systems have long treated semantic similarity as a proxy for relevance.
For constraint-sensitive queries, this proxy can fail when a document is topically close to the query but supports the opposite constraint direction, such as satisfying an attribute that should be excluded or affirming a relation that should be negated.
We study this failure as constraint-violating evidence exposure and propose CoDeR, a local constraint-compatible dense retrieval method that separates topical relevance from constraint compatibility.
CoDeR keeps a standard topical encoder for candidate coverage and adds a compatibility scorer, implemented as a bi-encoder, trained with lexical-polarity supervision over contrastive satisfying and violating evidences.
The compatibility signal can be used to rescore topical candidates or to retrieve an auxiliary compatibility-oriented candidate set, producing a ranked document list without external Large Language Model~(LLM) calls at inference time.
We evaluate CoDeR on controlled diagnostics and public negative-constraint retrieval benchmarks.
Across three controlled diagnostic sets targeting antonymy, negation, and exclusion, CoDeR reduces V@2 by 20.59, 23.53, and 5.77 points relative to the strongest non-CoDeR baselines, and improves FVR by pushing the first violating document deeper in the ranking.
Our source code and datasets are available at {\color{blue}https://github.com/NICE-HKU/CoDeR}.
\end{abstract}

\section{Introduction}
Information retrieval (IR) systems are increasingly used in real-world applications where users must locate evidence from large, domain-specific, or corpus-specific document collections.
In such systems, risk begins at retrieval because the retriever decides which documents are exposed to users, readers, or downstream models~\cite{lewis2020retrieval}.
Existing dense retrievers are optimized primarily for semantic similarity, where they reward documents that are close to the query in meaning~\citep{karpukhin2020dense, izacard2021unsupervised, xiao2024c} rather than documents that satisfy the intended constraint direction.
This design becomes fragile when the query contains explicit constraints, preferences, or exclusions.
For such constraint-sensitive queries, relevance is directional rather than merely topical.
A document may mention the same entities, attributes, and domain vocabulary as the query, yet describe the condition that the user wants to avoid.
Thus, the highest-ranked document can be semantically close while being constraint-incompatible, making it a plausible but harmful retrieval result.

\begin{figure}[t]
  \includegraphics[width=\columnwidth]{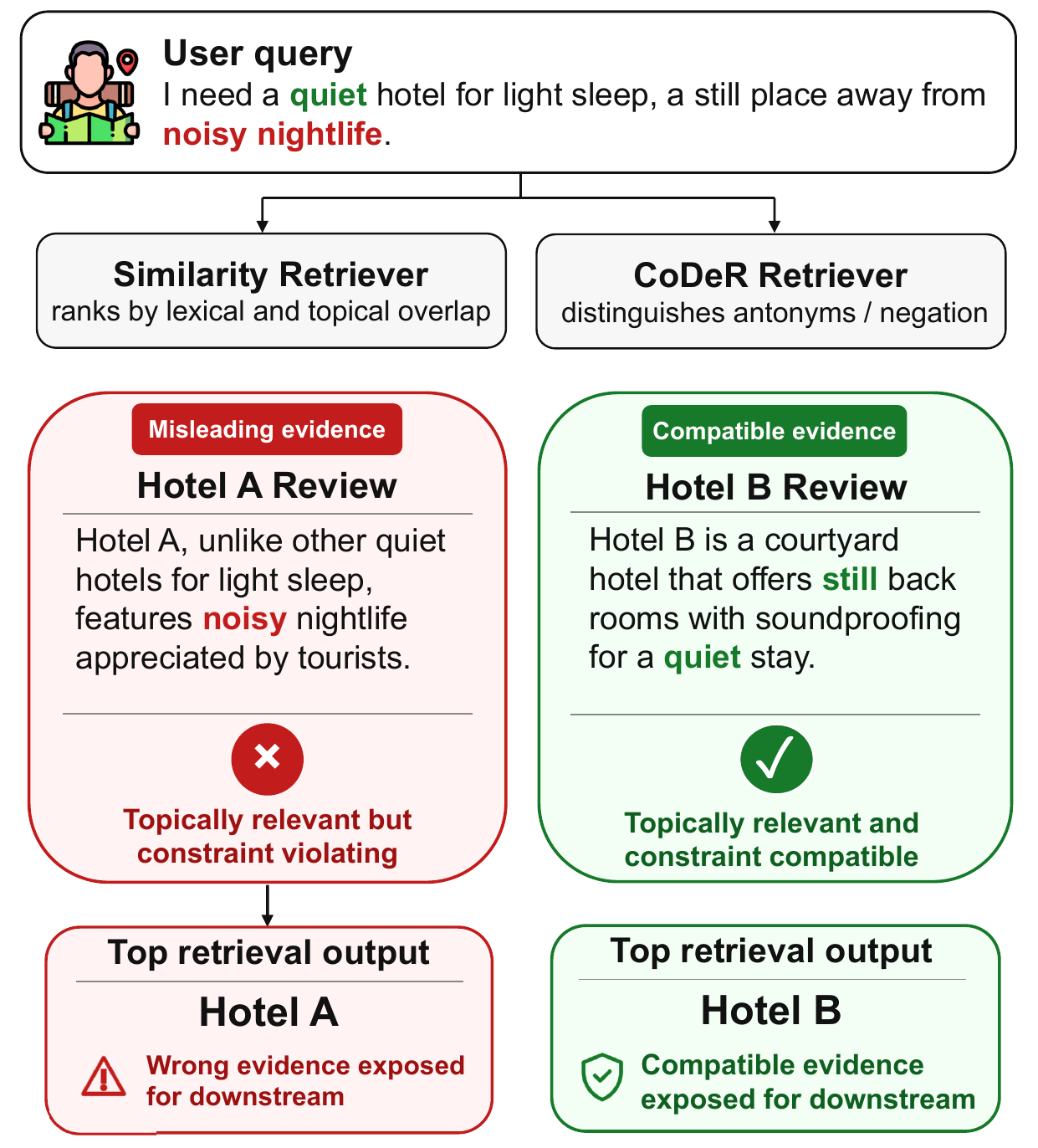}
  \captionsetup{font=small}
  \caption{Example of constraint-violating evidence exposure. A topically relevant review can repeat the query's excluded condition and therefore support the wrong constraint direction, even though it overlaps strongly with the query.}
  \label{fig:false_retrieve}
\end{figure}

This failure can arise across real-world retrieval settings where user intent is expressed through constraints, including medical retrieval with exclusion conditions, legal search, product recommendation, travel and hotel preferences, and enterprise knowledge-base question answering.
Figure~\ref{fig:false_retrieve} illustrates one such case with a hotel-search query that asks for a quiet stay while avoiding noisy nightlife. 
A standard retriever may rank a review of \textit{Hotel A} above a more suitable review of \textit{Hotel B} because it repeats the query's key terms, even though it describes the noise the user wants to avoid. 
By contrast, the \textit{Hotel B} review provides compatible cues such as quiet back rooms and soundproofing, but overlaps less directly with the query. 
We refer to such topically plausible but constraint-incompatible candidates as constraint-violating evidence.

This paper studies constraint-violating evidence exposure as a retrieval-side failure mode. 
We investigate whether a retriever can reduce early exposure to constraint-violating evidence while preserving topical coverage, and measure this risk with violation-oriented diagnostics such as V@k and FVR. 
Recent work addresses negative-constraint retrieval through logical reasoning pipelines or query-side embedding optimization~\cite{xu2025logical, lee2026deo}. 
Whether local retrieval can directly score document-level constraint compatibility remains underexplored.

We propose CoDeR, a local information retrieval method that decouples topical relevance from constraint compatibility. 
Constraint-sensitive retrieval must preserve topical coverage while suppressing documents that are topically plausible but constraint-incompatible. 
CoDeR retains a standard topical encoder for query-intent matching and introduces a compatibility encoder that scores whether a candidate document satisfies the constraint expressed in the query. 
Trained with lightweight lexical-polarity supervision over antonymy, negation, and exclusion, this encoder separates satisfying documents from nearby documents with the opposite constraint direction. 
A modular integration policy combines the topical and compatibility signals to produce a compatibility-aware list.
At inference time, CoDeR uses only local encoder scoring without external LLM calls.

We evaluate CoDeR on controlled diagnostics and public negative-constraint benchmarks. 
Across antonymy, negation, and exclusion diagnostics, CoDeR lowers V@2 from 73.53, 72.55, and 53.85 to 52.94, 49.02, and 48.08, respectively, while maintaining competitive topical retrieval quality. 
On NevIR and ExcluIR, CoDeR obtains the strongest violation-oriented performance, suggesting that local compatibility scoring can reduce early violation exposure relative to semantic-similarity-based retrieval.

Our main contributions are as follows:
\begin{itemize}
    \item We formulate constraint-violating evidence exposure as a retrieval-side failure mode for constraint-sensitive queries, and evaluate it with V@k and FVR diagnostics.
    
    \item We propose CoDeR, a local constraint-compatible retrieval method that composes topical retrieval with compatibility signals learned from lexical-polarity contrasts.
    
    \item We evaluate CoDeR on controlled diagnostic sets and public negative-constraint retrieval benchmarks, showing reduced early violation exposure while maintaining competitive topical retrieval quality.
\end{itemize}


\section{Related Work}

Dense retrievers encode queries and documents into a shared embedding space and rank documents by lexical or semantic similarity~\cite{karpukhin2020dense, izacard2021unsupervised, xiong2020approximate}. 
This paradigm is effective for topical matching, but it can fail when the user's intent depends on polarity, negation, or exclusion~\cite{weller2024nevir, zhang2025excluir}. 
Recent work has therefore explored retrieval methods that modify the query, document representation, or scoring procedure to better capture complex intent~\cite{gao2023precise}. 
CoDeR follows this embedding-side retrieval perspective, but focuses on separating topical relevance from constraint compatibility.

The closest line of work studies negative-constraint and negation-aware retrieval.
NS-IR introduces NegConstraint and reranks candidates through logical consistency after translating queries and documents into first order logic \citep{xu2025logical}. 
DEO decomposes negation aware queries into positive and negative components and directly optimizes query embeddings without training \citep{lee2026deo}. 
Both methods address important limitations of standard neural retrievers, but differ from CoDeR in deployment path and scoring target. 
Logic based pipelines often require translation or judging modules that increase latency and deployment complexity when implemented with strong external LLMs, while query embedding optimization is lightweight but mainly changes the query representation. 
CoDeR instead learns a local constraint compatibility scorer that directly evaluates candidate documents and produces a compatibility-aware ranked list without external LLM calls at inference time.

Under lexical polarity, both sparse and dense retrievers can become unreliable because they reward lexical overlap or semantic similarity.
Antonymy, negation, and exclusion can make satisfying and violating documents topically similar, turning constraint violations into structured hard negatives rather than ordinary irrelevant passages~\citep{wasserman2025docrerank,cho2026rare}. 
Prior work uses hard negatives to expose retrieval failures or improve reranker training~\cite{qu2021rocketqa, wasserman2025docrerank}. 
CoDeR instead targets hard negatives from constraint violation and uses lightweight lexical polarity supervision to separate satisfying evidence from opposite-direction evidence.
\section{Problem Formulation}

Given a query $q$ and a document corpus $\mathcal{D}=\{d_i\}_{i=1}^{N}$, a retriever assigns each $d\in\mathcal{D}$ a score and returns the top $k$ documents as a ranked list
\begin{equation}
    \label{eq:retrieved_list}
    R_k(q)=[r_1,\ldots,r_k].
\end{equation}
Standard retrievers typically rank documents by lexical overlap or semantic similarity. This is insufficient for constraint-sensitive queries, where a document can be topically related to the query while violating an explicit user requirement.

For analysis, we view a constraint-sensitive query as containing a topical component $t_q$ and a constraint component $c_q$, written as $q=(t_q,c_q)$. This decomposition is only used to define the task and does not assume an explicit symbolic parser at inference time. For each query $q$, documents are assigned query-dependent evidence labels. A satisfying document $d\in S_q$ matches the topic $t_q$ and satisfies the constraint $c_q$. A violating document $d\in V_q$ matches the topic but violates the constraint. Neutral topical documents are related to $t_q$ but do not explicitly satisfy or violate $c_q$. The same document may therefore be satisfying for one query and violating for another.

A constraint-compatible retriever should preserve topical coverage while ranking satisfying evidence above topically similar violating evidence. We express this desideratum as
\begin{equation}
    \label{eq:constraint_compatible_objective}
    f(q,d^+) > f(q,d^-), \quad d^+\in S_q,\ d^-\in V_q,
\end{equation}
where $f$ denotes the final retrieval score.

We evaluate whether violating evidence enters the retrieved context using retrieval-side diagnostics. Let $\bar{R}_k(q)=\{r_j:1\leq j\leq k\}$ be the set of returned documents. The violation indicator at depth $k$ is
\begin{equation}
    \label{eq:violation_at_k}
    \mathrm{V@k}(q)=\mathbb{I}\left[\bar{R}_k(q)\cap V_q\neq\emptyset\right].
\end{equation}
We also measure the first violating rank. Let
$J_k(q)=\{j\leq k \mid r_j\in V_q\}\cup\{k+1\}$.
\begin{equation}
    \label{eq:first_violating_rank}
    \mathrm{FVR}_k(q)=\min J_k(q).
\end{equation}
These metrics are retrieval-side risk indicators. 
They do not by themselves determine downstream answer behavior, but they measure whether the top-ranked retrieval results contain documents that conflict with the user's stated constraint.

\section{CoDeR: Constraint-Compatible Retrieval}
\label{sec_CoDeR}
CoDeR instantiates the constraint-compatible retrieval objective in Eq.~\ref{eq:constraint_compatible_objective} as a local information retrieval method.
A topical encoder $E_T$ produces a topical relevance score $s_T(q,d)$, while a constraint compatibility encoder $E_C$ produces a compatibility score $s_C(q,d)$.
An integration policy combines these scores to produce a compatibility-aware ranked document list.
CoDeR operates before any optional downstream reranker or generator and is evaluated by the ranked list it directly returns.

\subsection{Constraint Compatibility Encoder}

The constraint compatibility encoder $E_C$ estimates whether a document satisfies the constraint expressed in the query.
Unlike a topical retriever, which mainly rewards semantic closeness to the query, $E_C$ is trained to distinguish constraint-satisfying evidence from topically similar constraint-violating evidence.
In our implementation, $E_C$ is a bi-encoder, so the query and document are encoded independently~\cite{reimers2019sentence, karpukhin2020dense}.
Given a query $q$ and a document $d$, it computes a compatibility score by vector similarity,
\begin{equation}
    \label{eq:constraint_compatibility_score}
    s_C(q,d)=\mathrm{sim}\left(E_C(q),E_C(d)\right).
\end{equation}
where $\operatorname{sim}(\cdot,\cdot)$ denotes cosine similarity between the two embeddings.
A higher value of $s_C(q,d)$ indicates stronger evidence that $d$ satisfies the requirement expressed in $q$.

\begin{figure}[t]
    \centering
    \includegraphics[width=\columnwidth]{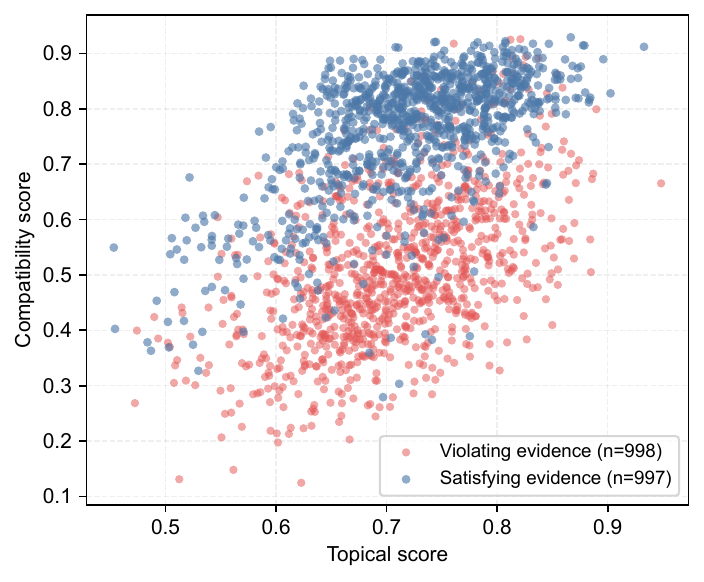}
    \captionsetup{font=small}
    \caption{
    Topicality versus compatibility on ExcluIR.
    Each point is a labeled candidate from the topical encoder's top-$k$ results.
    Satisfying and violating evidence overlap along the topical-score axis but separate along the compatibility-score axis, showing that constraint violations are often topical near neighbors rather than off-topic noise.
    }
    \label{fig:topical_compatibility_scatter}
\end{figure}

Because $E_C$ is a bi-encoder, document embeddings can be precomputed and indexed for two uses in CoDeR.
In the \textbf{sequential policy}, $E_C$ supplies a compatibility score over the topical candidate set $\mathcal{C}_T(q)$.
In the \textbf{union policy}, $E_C$ retrieves a compatibility-driven candidate set $\mathcal{C}_C(q)$ from the same corpus, allowing constraint-compatible evidence that may be missed by the topical retriever to enter the candidate pool.
Thus, $E_C$ complements topical retrieval with a constraint-compatibility signal, helping separate satisfying evidence from topically similar violations.
Implementation details, including the base encoder, query prefix, and training data split, are provided in Section~\ref{subsec:experimental_setup} and Appendix~\ref{app:constraint_encoder_training}.
Figure~\ref{fig:topical_compatibility_scatter} illustrates why this separate compatibility signal is needed.
On ExcluIR, satisfying and violating evidence occupy overlapping high-topicality regions under the topical encoder, indicating that violations are not merely off-topic outliers.
The trained compatibility encoder separates them along the compatibility axis, motivating CoDeR's use of topical scoring for coverage and compatibility scoring for constraint direction.

\subsection{Lexical Polarity Supervision}
\label{subsec:lexical_polarity_supervision}

We train the constraint compatibility encoder $E_C$ with lightweight lexical polarity supervision.
The supervision is constructed from antonymy, negation, and exclusion patterns that change the direction of a user constraint while preserving the surrounding topic.
For each query $q$, we construct a satisfying document $d^+$ and a violating document $d^-$.
Both documents remain topically related to the query which discourages the model from solving the task by topical matching alone.
This design concentrates the learning signal on constraint compatibility rather than general topical relevance.

We optimize $E_C$ with a multiple negatives ranking objective over triples $(q,d^+,d^-)$~\cite{oord2018representation}.
The satisfying document is treated as the positive document, while the violating document serves as an explicit hard negative. Other documents in the same batch are also used as in-batch negatives. Let $\mathcal{N}(q)$ denote the negative set for query $q$, including the explicit violating document and in-batch negatives. The training loss is
\begin{equation}
    \label{eq:constraint_encoder_loss}
    \begin{aligned}
        \mathcal{L}
        &=
        -\beta s_C(q,d^+)
        +
        \log Z(q),\\
        Z(q)
        &=
        \exp\left(\beta s_C(q,d^+)\right)\\
        &\quad+
        \sum_{d^- \in \mathcal{N}(q)}
        \exp\left(\beta s_C(q,d^-)\right).
    \end{aligned}
\end{equation}
where $\beta$ is a scaling factor. This objective encourages $E_C$ to assign higher compatibility scores to satisfying evidence than to topically similar evidence with the wrong constraint direction.

\subsection{Modular Integration of CoDeR}
\label{subsec_modular_integration}

An integration policy converts the topical and compatibility scores into the final ranked document list. 
Let $s_T(q,d)$ denote the topical relevance score from $E_T$, and let $s_C(q,d)$ denote the constraint compatibility score from $E_C$. 
The policy may vary in how candidates are generated, how the two scores are combined, and how low-compatibility candidates are filtered. 
This modular design is not tied to a single implementation and can support different candidate generation, scoring, and filtering strategies.
In this paper, we instantiate this design with two variants, CoDeR-Seq and CoDeR-Union. 
CoDeR-Seq is a sequential compatibility-scoring policy that keeps the standard topical encoder as the candidate generator and uses $E_C$ to rescore and filter topical candidates.
CoDeR-Union is a union candidate policy that lets both $E_T$ and $E_C$ retrieve candidates, then merges the two candidate sets before fused ranking. 
Thus, the two variants share the same topical and compatibility encoders, but differ in whether the compatibility encoder only scores topical candidates or also retrieves additional candidates from the corpus.

\begin{figure}[t]
    \centering
    \includegraphics[width=\columnwidth]{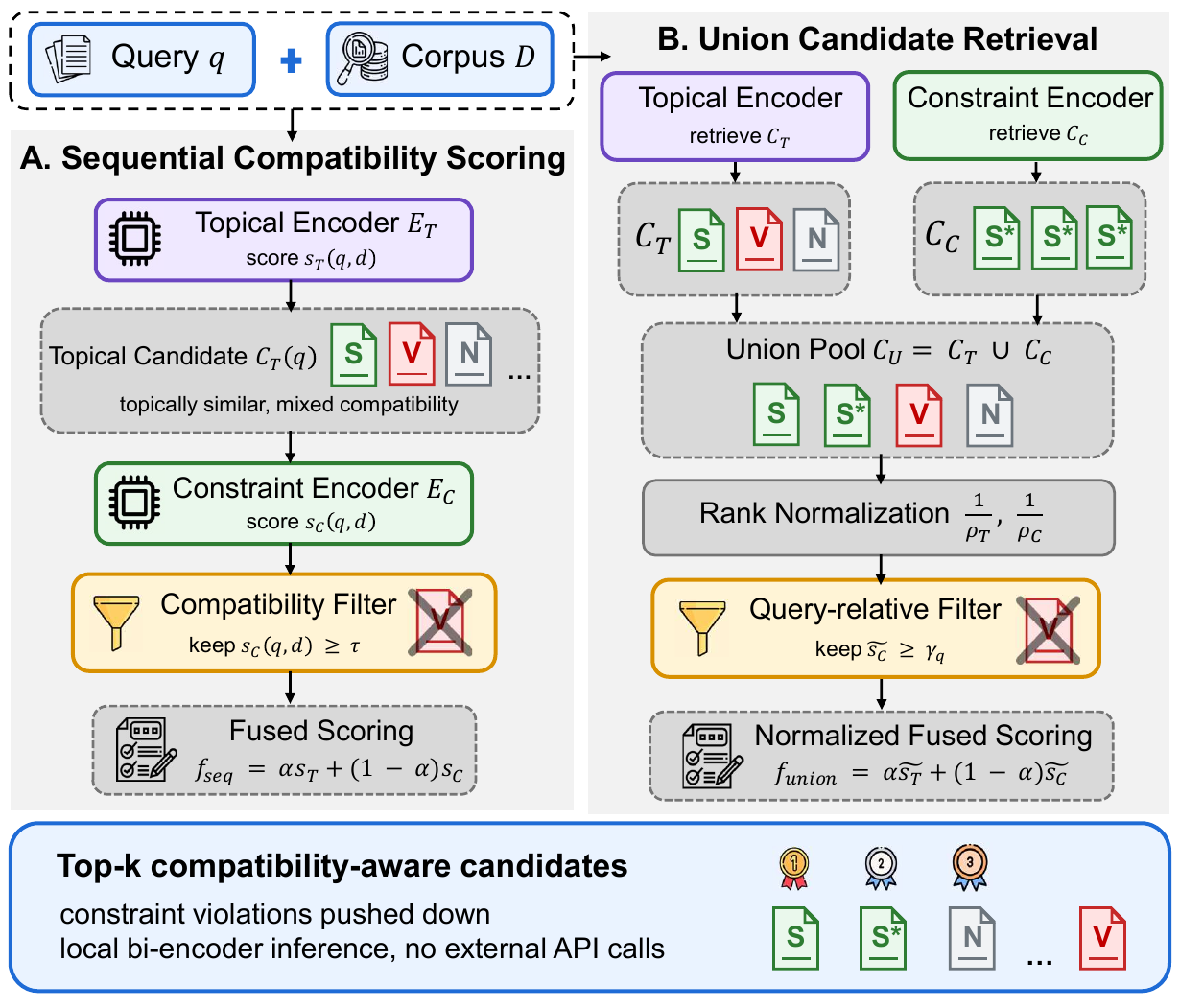}
    \captionsetup{font=small}
    \caption{
    Overview of the two CoDeR integration policies.
    (A) Sequential Integration scores topical candidates with $E_C$.
    (B) Union Candidate Integration merges candidates retrieved by $E_T$ and $E_C$ before compatibility-aware ranking.
    }
    \label{fig_coder_frameworks}
\end{figure}

\paragraph{Sequential Integration}
Sequential Integration is a lightweight compatibility-scoring policy shown in Fig.~\ref{fig_coder_frameworks}(A).
It first retrieves a topical candidate set $\mathcal{C}_T(q)$ using $E_T$, and then scores each candidate with $E_C$.
Candidates whose compatibility scores fall below a threshold $\tau$ are filtered:
\begin{equation}
    \label{eq:sequential_filter}
    \mathcal{C}_{\mathrm{seq}}(q)
    =
    \{d \in \mathcal{C}_T(q) : s_C(q,d) \geq \tau\}.
\end{equation}
We rank the remaining candidates by:
\begin{equation}
    \label{eq:sequential_fusion_score}
    f_{\mathrm{seq}}(q,d)
    =
    \alpha s_T(q,d)
    +
    (1-\alpha)s_C(q,d).
\end{equation}
where $\tau$ is the compatibility threshold, and $\alpha \in [0,1]$ controls the balance between topical relevance and constraint compatibility.
This policy requires one topical retrieval call and local compatibility scoring over $\mathcal{C}_T(q)$.
It allows an existing retrieval pipeline to add compatibility-aware reranking without changing the original candidate generation stage.

\paragraph{Union Candidate Integration}
Union Candidate Integration, shown in Fig.~\ref{fig_coder_frameworks}(B), relaxes the coverage bottleneck of the sequential policy by allowing both encoders to participate in candidate retrieval.
The topical encoder retrieves $\mathcal{C}_T(q)$, while the constraint compatibility encoder retrieves candidate set $\mathcal{C}_C(q)$.
The final candidate pool is their union:
\begin{equation}
    \label{eq:union_candidate_pool}
    \mathcal{C}_U(q)
    =
    \mathcal{C}_T(q)
    \cup
    \mathcal{C}_C(q).
\end{equation}

After forming $\mathcal{C}_U(q)$, CoDeR rescores every document in the union pool with both encoders.
Because $s_T$ and $s_C$ may have different score scales, we use rank-based normalized scores.
Let $\rho_T(q,d)$ and $\rho_C(q,d)$ denote the rank positions of $d$ inside $\mathcal{C}_U(q)$ when candidates are sorted by $s_T$ and $s_C$, respectively.
We define
\begin{equation}
    \label{eq:rank_normalization}
    \begin{aligned}
        \widetilde{s}_T(q,d) &= \frac{1}{\rho_T(q,d)},\\
        \widetilde{s}_C(q,d) &= \frac{1}{\rho_C(q,d)}.
    \end{aligned}
\end{equation}
The normalized scores are then combined:
\begin{equation}
    \label{eq:union_fusion_score}
    f_{\mathrm{union}}(q,d)
    =
    \alpha \widetilde{s}_T(q,d)
    +
    (1-\alpha)\widetilde{s}_C(q,d).
\end{equation}

The union policy applies a query-relative compatibility filter.
Given a query-specific threshold $\gamma_q$, it retains candidates with $\widetilde{s}_C(q,d) \geq \gamma_q$ and returns the top $k$ documents ranked by Eq.~\ref{eq:union_fusion_score}.
If no candidate passes the filter, CoDeR skips only the hard cutoff and ranks the full union pool by the same fused score, preserving compatibility as an active ranking signal.


CoDeR returns a ranked document list and does not modify any downstream reader, reranker, or generator. 
This separation allows the retrieval method to be evaluated directly. 
Optional downstream components can be attached after CoDeR, while the main experiments isolate the embedding-side compatibility signal by evaluating the direct ranked output.

\section{Experiments}
We evaluate CoDeR from three retrieval-centered perspectives: preservation of ordinary topical retrieval quality, reduction of early violation exposure, and evaluation on public negative-constraint benchmarks.
We also report inference-time overhead and API usage.
A small downstream probe is reported in Appendix~\ref{app:end2end} as a preliminary end-to-end validation.

\subsection{Experimental Setup}
\label{subsec:experimental_setup}

\paragraph{Datasets and metrics.}
We use BEIR-style datasets~\cite{thakur2021beir} for topical-retrieval preservation, controlled antonymy/negation/exclusion diagnostics for constraint violations, public negative-constraint benchmarks for transfer.
For ordinary retrieval preservation, we report nDCG and MAP.
For constraint-aware retrieval, we report V@$k$ and FVR: V@$k$ measures whether at least one constraint-violating document appears in the top $k$ results, while FVR measures the rank of the first violation.
Lower V@$k$ and higher FVR are better.
FVR captures the earliest rank at which incompatible evidence appears, instead of depending on a single fixed top-$k$ threshold~\cite{liu2024lost}.

\paragraph{Baselines.}
We group the comparison methods by retrieval mechanism.
The lexical retrieval baseline is BM25, which represents sparse term-matching retrieval~\cite{robertson2009probabilistic}.
Dense semantic retrieval baselines include BGE and Contriever, which rank documents mainly by learned semantic similarity.
The query-generation baseline is HyDE, which retrieves using LLM-generated hypothetical documents.
Constraint-oriented baselines include NS-IR and DEO, the methods closest to our setting: NS-IR targets constraint-aware retrieval, while DEO is designed for exclusion-oriented negative-constraint retrieval.
We report two proposed compatibility-aware retrieval variants, CoDeR-Seq and CoDeR-Union, corresponding to the sequential and union policies in Section~\ref{subsec_modular_integration}.
All methods are evaluated by their direct retrieval output without an additional cross-encoder reranker.
This setting follows embedding-side retrieval comparisons and isolates whether the retrieval method itself produces a more constraint-compatible ranked list.

\paragraph{Implementation details.}
The constraint compatibility encoder $E_C$ is initialized from \texttt{bge-large-en-v1.5} and trained as a bi-encoder.
Training has two stages, using WordNet-derived word-level lexical polarity triplets and sentence-level triplets from 800 NevIR and 800 ExcluIR training queries excluded from testing.
All queries, triplets, and associated documents used to train $E_C$ are removed from evaluation splits and reported test results.
We use the same query prefix for training and inference.
All reported results are averaged over 10 independent runs, with both the method ranking and CoDeR gains remain stable across runs.

\subsection{General Retrieval Quality}
\label{subsec:general_retrieval_quality}

\begin{table*}[t]
    \small
    \centering
    \resizebox{\textwidth}{!}{
    \begin{tabular}{l|cc|cc|cc|cc|cc|cc}
        \toprule
        \multirow{2.5}{*}{Methods}
        & \multicolumn{2}{c|}{SciFact}
        & \multicolumn{2}{c|}{ArguAna}
        & \multicolumn{2}{c|}{FiQA}
        & \multicolumn{2}{c|}{NFCorpus}
        & \multicolumn{2}{c|}{\makecell{CQA\\Android}}
        & \multicolumn{2}{c}{SciDocs} \\
        \cmidrule{2-3}\cmidrule{4-5}\cmidrule{6-7}\cmidrule{8-9}\cmidrule{10-11}\cmidrule{12-13}
        & nDCG$\uparrow$ & MAP$\uparrow$ & nDCG$\uparrow$ & MAP$\uparrow$ & nDCG$\uparrow$ & MAP$\uparrow$ & nDCG$\uparrow$ & MAP$\uparrow$ & nDCG$\uparrow$ & MAP$\uparrow$ & nDCG$\uparrow$ & MAP$\uparrow$ \\
        \midrule
        BM25
        & 64.82 & 60.37
        & 34.40 & 22.56 
        & 23.25 & 17.20 
        & 30.98 & 22.36 
        & 39.08 & 34.32 
        & 15.04 & 8.76 \\

        BGE
        & 74.77 & 69.93 
        & 45.99 & 32.25 
        & 45.00 & 37.15 
        & \textbf{38.15} & \textbf{28.76}
        & \textbf{50.22} & \textbf{43.89}
        & \textbf{22.63} & \textbf{13.80} \\

        Contriever 
        & 55.04 & 49.43 
        & 33.55 & 21.90 
        & 12.41 & 8.81 
        & 27.12 & 18.62 
        & 30.37 & 25.56 
        & 10.97 & 5.85 \\
        \specialrule{0.25pt}{2pt}{0pt}

        HyDE
        & 67.44 & 61.98 
        & 30.41 & 19.28 
        & 28.51 & 22.16 
        & 31.59 & 22.46 
        & 21.55 & 17.56 
        & 14.21 & 7.95 \\

        NS-IR
        & 74.44 & 69.92 
        & 45.35 & 32.31 
        & 42.68 & 34.80 
        & 36.92 & 27.66 
        & 49.94 & 43.61 
        & 22.59 & 13.42 \\

        DEO 
        & 72.09 & 67.19 
        & \textbf{70.27} & \textbf{63.08}
        & 43.62 & 35.56 
        & 36.68 & 14.27 
        & 44.59 & 37.91 
        & 22.33 & 13.47 \\
        
        \specialrule{0.25pt}{2pt}{0pt}

        CoDeR-Seq
        & 74.77 & 69.93 
        & 49.36 & 35.40 
        & \textbf{45.00} & \textbf{37.15} 
        & 37.24 & 28.00 
        & 48.02 & 42.15 
        & 21.42 & 12.86 \\

        CoDeR-Union 
        & \textbf{74.91} & \textbf{70.17}
        & 48.78 & 35.00 
        & 44.51 & 36.80 
        & 36.36 & 27.12 
        & 48.17 & 42.22 
        & 21.60 & 12.96 \\

        \bottomrule
    \end{tabular}
    }
    \caption{
General retrieval quality on BEIR-style benchmarks.
We report nDCG@10 and MAP@10.
These datasets do not explicitly target constraint violations and are used to test whether CoDeR preserves topical retrieval quality.
}
    \label{tab:beir}
\end{table*}

Although CoDeR targets constraint-sensitive retrieval, an information retrieval method must still preserve ordinary topical coverage.
Table~\ref{tab:beir} therefore serves as a preservation check rather than a claim of general-purpose retrieval superiority.
Because these datasets mostly reward topical matching, the strongest dense retrievers can remain competitive without modeling constraint direction.
The relevant observation is that adding $E_C$ keeps CoDeR close to the base dense retriever while introducing a compatibility signal that becomes important when topical relevance and constraint satisfaction diverge.

\subsection{Constraint-Aware Retrieval on Diagnostic Datasets}
\label{subsec:diagnostic_retrieval}

As general retrieval benchmarks do not test whether retrievers separate topical relevance from constraint compatibility, we evaluate on controlled antonymy, negation, and exclusion datasets where satisfying and violating documents are topical near neighbors with opposite constraint directions.
Table~\ref{tab:selfbuilt} reports violation-oriented metrics for this on-topic but constraint-unsafe setting, with construction details and examples in Appendix~\ref{app:dataset_construction}.

\begin{table*}[t]
    \small
    \centering
    \resizebox{\textwidth}{!}{
    \begin{tabular}{l|ccccc|ccccc|ccccc}
        \toprule
        \multirow{2.5}{*}{Methods}
        & \multicolumn{5}{c|}{Antonym}
        & \multicolumn{5}{c|}{Negation}
        & \multicolumn{5}{c}{Exclusion} \\
        \cmidrule{2-6}\cmidrule{7-11}\cmidrule{12-16}
        & V@2$\downarrow$ & V@3$\downarrow$ & V@5$\downarrow$ & V@10$\downarrow$ & FVR$\uparrow$
        & V@2$\downarrow$ & V@3$\downarrow$ & V@5$\downarrow$ & V@10$\downarrow$ & FVR$\uparrow$
        & V@2$\downarrow$ & V@3$\downarrow$ & V@5$\downarrow$ & V@10$\downarrow$ & FVR$\uparrow$ \\
        \midrule
        BM25
        & 90.20 & 98.04 & 100.00 & 100.00 & 2.11
        & 98.03 & 100.00 & 100.00 & 100.00 & 2.00
        & 76.47 & 87.25 & 97.06 & 100.00 & 2.26 \\
        
        BGE
        & 81.37 & 92.16 & 99.02 & 100.00 & 2.30
        & 90.20 & 96.08 & 100.00 & 100.00 & 2.02
        & 69.61 & 82.35 & 88.24 & 95.10 & 2.93 \\
        
        Contriever 
        & 78.43 & 95.10 & 100.00 & 100.00 & 2.21
        & 83.33 & 92.16 & 99.02 & 100.00 & 2.25
        & 77.45 & 89.22 & 96.08 & 100.00 & 2.20 \\
        \specialrule{0.25pt}{2pt}{0pt}
        
        HyDE
        & 82.35 & 92.16 & 100.00 & 100.00 & 2.08
        & 84.31 & 91.18 & 99.02 & 100.00 & 2.06
        & 66.67 & 83.33 & 96.08 & 99.02 & 2.48 \\
        
        NS-IR
        & 75.49 & 92.16 & 99.02 & 100.00 & 2.34
        & 84.31 & 97.06 & 100.00 & 100.00 & 2.15
        & 68.63 & 81.37 & 92.16 & 99.02 & 2.72 \\
        
        DEO        
        & 73.53 & 85.29 & 98.04 & 100.00 & 2.42
        & 72.55 & 86.27 & 94.12 & 97.06 & 2.68
        & 53.85 & 71.15 & 79.81 & 88.46 & 3.76 \\
        
        \specialrule{0.25pt}{2pt}{0pt}
        
        CoDeR-Seq
        & 59.80 & \textbf{76.47} & \textbf{91.18} & 99.02 & 2.95
        & 50.98 & 73.53 & \textbf{84.31} & 92.16 & \textbf{3.53}
        & 50.00 & \textbf{60.58} & \textbf{74.04} & 84.62 & 4.26 \\
        
        CoDeR-Union 
        & \textbf{52.94} & \textbf{76.47} & 93.13 & \textbf{98.04} & \textbf{3.00}
        & \textbf{49.02} & \textbf{72.55} & 89.22 & \textbf{90.20} & 3.47
        & \textbf{48.08} & 61.54 & 75.96 & \textbf{81.73} & \textbf{4.37} \\
        
        \bottomrule
    \end{tabular}
    }
    \caption{
    Constraint-aware retrieval results on self-constructed Antonym, Negation, and Exclusion datasets.
    Metrics are defined in Section~\ref{subsec:experimental_setup}.
    }
    \label{tab:selfbuilt}
\end{table*}

The diagnostic results expose a progression in how different retrieval mechanisms handle constraints: BM25, BGE, and Contriever represent the basic retrieval regime, where lexical overlap or dense semantic similarity can find documents about the same entities and attributes but has no explicit reason to treat satisfying and violating counterparts as opposite retrieval outcomes.
This makes the diagnostic setting difficult as the violating document is not off-topic noise, but a topical near neighbor pointing in the wrong constraint direction.

HyDE and NS-IR make queries more informative through LLM-based expansion, rewriting, or constraint reasoning, but the inherited topical embedding space still limits their ability to separate satisfying and violating evidence consistently.

DEO and CoDeR reflect a stronger intervention by putting pressure on constraint-violating evidence itself, treating it as something to be separated or penalized at the embedding and scoring level rather than only rephrasing the query.
CoDeR makes this idea explicit by training $E_C$ with satisfying--violating contrasts and applying the resulting compatibility signal over topically plausible candidates.
Thus, the early V@$k$ and FVR gains suggest that constraint-aware retrieval benefits from a representation where constraint direction is separable from topical relatedness, rather than from query rewriting alone.

\subsection{Evaluation on Published Negative-Constraint Benchmarks}
\label{subsec:published_benchmarks}

We next evaluate CoDeR on published negative-constraint retrieval benchmarks, including NevIR, ExcluIR, and NegConstraint.
NevIR focuses on negation-aware retrieval, while ExcluIR evaluates explicit exclusion constraints.
For NevIR and ExcluIR, we map paired opposite-direction evidence to the violating set and report V@$k$ and FVR.
NegConstraint follows a different released protocol, so we report its available metrics separately.
All queries used to train $E_C$ are removed from the corresponding NevIR and ExcluIR evaluation splits, yielding held-out evaluations within the published negative-constraint benchmarks.

\begin{table*}[h]
    \small
    \centering
    \resizebox{\textwidth}{!}{
    \begin{tabular}{l|ccccc|ccccc}
        \toprule
        \multirow{2.5}{*}{Methods}
        & \multicolumn{5}{c|}{NevIR}
        & \multicolumn{5}{c}{ExcluIR} \\
        \cmidrule{2-6}\cmidrule{7-11}
        & V@2$\downarrow$ & V@3$\downarrow$ & V@5$\downarrow$ & V@10$\downarrow$ & FVR$\uparrow$
        & V@2$\downarrow$ & V@3$\downarrow$ & V@5$\downarrow$ & V@10$\downarrow$ & FVR$\uparrow$ \\
        \midrule
        BM25
        & 43.00 & 65.20 & 85.00 & 90.40 & 3.56
        & 90.20 & 93.90 & 96.60 & 97.70 & 1.94 \\
        
        BGE
        & 36.80 & 60.30 & 85.60 & 90.30 & 3.70
        & 96.10 & 97.70 & 98.20 & 99.10 & 1.75 \\
        
        Contriever 
        & 42.10 & 64.40 & 84.20 & 88.50 & 3.69
        & 91.60 & 93.90 & 96.20 & 98.00 & 1.91 \\
        \specialrule{0.25pt}{2pt}{0pt}
        
        HyDE
        & 41.30 & 63.40 & 86.10 & 90.00 & 3.58
        & 57.40 & 66.00 & 75.00 & 81.80 & 4.21 \\
        
        NS-IR
        & 35.80 & 60.90 & 84.80 & 91.40 & 3.71
        & 95.60 & 97.10 & 98.10 & 98.70 & 1.79 \\
        
        DEO        
        & 31.10 & 56.50 & 81.90 & 88.60 & 4.02
        & 51.10 & 59.60 & 66.10 & 74.30 & 4.93 \\
        
        \specialrule{0.25pt}{2pt}{0pt}
        
        CoDeR-Seq
        & \textbf{28.50} & \textbf{52.90} & 79.40 & 86.20 & 4.26
        & 47.70 & \textbf{56.60} & \textbf{64.40} & 73.70 & 5.06 \\
        
        CoDeR-Union
        & 30.20 & 53.20 & \textbf{78.80} & \textbf{86.00} & \textbf{4.28}
        & \textbf{45.00} & 57.20 & 65.30 & \textbf{65.70} & \textbf{5.33} \\
        
        \bottomrule
    \end{tabular}
    }
    \caption{
Results on published NevIR and ExcluIR benchmarks.
Metrics are defined in Section~\ref{subsec:experimental_setup}.
}
    \label{tab:benchmarks}
\end{table*}

Table~\ref{tab:benchmarks} shows that the compatibility-scoring mechanism remains effective on NevIR and ExcluIR beyond our diagnostic construction.
Across both negation and exclusion, topical-matching or query-rewriting methods still expose violations early, while methods that score or penalize constraint-violating candidates move the first violation deeper, suggesting that $E_C$ captures a compatibility signal beyond a single diagnostic template.

Table~\ref{tab:negconstraint} reports results on NegConstraint under its released metrics, since the full label structure needed for our violation diagnostics is unavailable.
The gains over NS-IR provide a complementary check: even under standard ranking metrics, local compatibility scoring improves retrieval under negative constraints.
Together with the violation-oriented results above, this suggests that the benefit comes from how candidate evidence is represented and scored, not simply from better prompt wording.

\subsection{Ablation Study Summary}
We summarize the component ablations here and provide full results in Appendix~\ref{app:ablation_study}. 
The ablations test whether CoDeR's gains come from the proposed separation between topical relevance and constraint compatibility, rather than from a stronger topical retriever, another dense encoder, or a post-hoc reranker. 
First, changing the topical encoder has a larger effect on ordinary retrieval quality than on the overall violation-control pattern when the compatibility encoder is fixed. 
Second, replacing the compatibility encoder with generic dense encoders weakens violation-oriented behavior, suggesting semantic similarity alone does not provide the same constraint-compatibility signal. 
Third, an additional reranker ablation shows that a standard reranker does not consistently delay the first violating document when topically similar violations have already entered the candidate set.
Together, these results support CoDeR's two-signal design, where topical retrieval preserves coverage and compatibility scoring controls constraint direction.

\subsection{Hyperparameter Policy Sensitivity}
\label{subsec:policy_sensitivity}

We briefly analyze policy sensitivity to verify that CoDeR is not driven by a single tuned hyperparameter point; full heatmaps and detailed analysis are provided in Appendix~\ref{app:policy_sensitivity}.
Across Negation and ExcluIR, the main trade-off is between preserving satisfying evidence, measured by Recall@10, and delaying the first violating document, measured by FVR.
CoDeR-Seq is more sensitive to the absolute threshold $\tau$ because it depends on raw compatibility-score calibration, whereas CoDeR-Union uses a relative filter percentile, $\gamma_q$, and shows smoother behavior across datasets.
The heatmaps confirm that $\alpha$ governs the trade-off between topical coverage and compatibility-driven ranking, and that filtering strength adjusts constraint enforcement across a continuous range rather than collapsing to a single tuned operating point.

\subsection{Efficiency Analysis}
\label{subsec:efficiency_analysis}

We report inference-time efficiency separately because time, external API cost, and token usage reflect deployment overhead rather than retrieval accuracy.
Table~\ref{tab:efficiency} aggregates measurements over Antonym, Negation, Exclusion, NevIR, ExcluIR, and NegConstraint, with full per-setting measurements in Appendix~\ref{app:full_efficiency_accounting}.
The reported time includes offline preparation such as model loading, corpus encoding, index construction, and cache initialization, plus online processing over all queries.

CoDeR runs locally after corpus pre-encoding where inference only requires query encoding, vector scoring, candidate filtering, and reranking, so it incurs no API tokens or external API cost.
By contrast, HyDE, NS-IR, and DEO repeatedly call LLM APIs for hypothetical document generation, logical translation, or query decomposition.
For fairness, those API-based baselines are evaluated with GPT-4o where applicable and their logged tokens and costs are included.

\begin{table}[t]
    \centering
    \small
    \begin{tabular}{l|ccc}
        \toprule
        Method & Time(h)$\downarrow$ & Cost(\$)$\downarrow$ & Tokens(K)$\downarrow$ \\
        \midrule
        HyDE & 46.67 & 32.30 & 3854.0 \\
        NS-IR & 53.76 & 32.22 & 4531.4 \\
        DEO & 9.64 & 5.12 & 1376.9 \\
        \specialrule{0.25pt}{2pt}{0pt}
        CoDeR-Seq & 0.08 & 0.00 & 0.0 \\
        CoDeR-Union & 0.07 & 0.00 & 0.0 \\
        \bottomrule
    \end{tabular}
    \caption{
    Aggregate inference efficiency across six constraint-aware retrieval settings. Time reports total wall-clock hours including offline preparation and online query execution. Cost and tokens report external API usage. Lower is better.
    }
    \label{tab:efficiency}
\end{table}

The efficiency results highlight the difference between prompt/API-based and representation-based interventions.
CoDeR amortizes constraint learning into $E_C$ and avoids repeated LLM calls at inference time, making it attractive for cost-sensitive and privacy-sensitive retrieval deployments.

\section{Conclusion}

We presented CoDeR, a local constraint-compatible information retrieval method for constraint-sensitive queries.
CoDeR targets cases where topically relevant documents support the wrong constraint direction, such as negation, exclusion, or antonymic preferences.
By combining topical relevance with a constraint compatibility scorer, CoDeR reduces early exposure to constraint-violating evidence on controlled diagnostics and public negative-constraint benchmarks, while largely preserving ordinary retrieval quality.
Our results suggest that constraint-sensitive retrieval should be evaluated not only by topical relevance, but also by whether early-ranked documents satisfy the direction of the user’s stated constraint.

\section*{Limitations}
CoDeR focuses on explicit negative constraints expressed through antonymy, negation, and exclusion. 
Other constraint types, such as numeric, temporal, multi-hop, implicit, or pragmatically ambiguous constraints, may require additional supervision or different compatibility signals. 
CoDeR is also trained with English lexical-polarity resources and sentence-level triplets from negative-constraint benchmarks. 
Although all training queries and associated documents are removed from the corresponding evaluation splits, the NevIR and ExcluIR results are best viewed as held-out in-domain evaluations rather than fully out-of-domain transfer. 
Applying CoDeR to new languages or specialized domains may require constructing polarity triplets and adapting the compatibility encoder.

Our main evaluation is retrieval-side. 
CoDeR aims to reduce early exposure to constraint-violating documents in the ranked list, but it does not by itself guarantee downstream answer factuality or safety. 
The small downstream probe is included only as a preliminary downstream validation.
Evaluating how compatibility-aware retrieval composes with downstream readers, rerankers, and generators remains future work.

Potential risks arise if compatibility scores are miscalibrated. 
CoDeR may incorrectly demote relevant documents when constraints are implicit, ambiguous, or expressed differently from the training patterns. 
In high-stakes settings such as medical or legal retrieval, CoDeR should therefore be used as a retrieval-side aid rather than a standalone safety mechanism, with downstream verification or human review when appropriate.

\bibliography{custom}

@inproceedings{wasserman2025docrerank,
  title={DocReRank: Single-page hard negative query generation for training multi-modal RAG rerankers},
  author={Wasserman, Navve and Heinimann, Oliver and Golbari, Yuval and Zimbalist, Tal and Schwartz, Eli and Irani, Michal},
  booktitle={Proceedings of the 2025 Conference on Empirical Methods in Natural Language Processing},
  pages={8651--8669},
  year={2025}
}

@inproceedings{xu2025logical,
  title={Logical Consistency is Vital: Neural-Symbolic Information Retrieval for Negative-Constraint Queries},
  author={Xu, Ganlin and Zhang, Zhoujia and Mei, Wangyi and Liang, Jiaqing and Lu, Weijia and Zhang, Xiaodong and Yang, Zhifei and Ma, Xiaofeng and Xiao, Yanghua and Yang, Deqing},
  booktitle={Findings of the Association for Computational Linguistics: ACL 2025},
  pages={1828--1847},
  year={2025}
}

@article{lee2026deo,
  title={DEO: Training-Free Direct Embedding Optimization for Negation-Aware Retrieval},
  author={Lee, Taegyeong and Park, Jiwon and Hwang, Seunghyun and Jang, JooYoung},
  journal={arXiv preprint arXiv:2603.09185},
  year={2026}
}

@article{cho2026rare,
  title={RARE: Redundancy-Aware Retrieval Evaluation Framework for High-Similarity Corpora},
  author={Cho, Hanjun and Lee, Jay-Yoon},
  journal={arXiv preprint arXiv:2604.19047},
  year={2026}
}

@inproceedings{karpukhin2020dense,
  title={Dense passage retrieval for open-domain question answering},
  author={Karpukhin, Vladimir and Oguz, Barlas and Min, Sewon and Lewis, Patrick and Wu, Ledell and Edunov, Sergey and Chen, Danqi and Yih, Wen-tau},
  booktitle={Proceedings of the 2020 conference on empirical methods in natural language processing (EMNLP)},
  pages={6769--6781},
  year={2020}
}

@article{izacard2021unsupervised,
  title={Unsupervised dense information retrieval with contrastive learning},
  author={Izacard, Gautier and Caron, Mathilde and Hosseini, Lucas and Riedel, Sebastian and Bojanowski, Piotr and Joulin, Armand and Grave, Edouard},
  journal={arXiv preprint arXiv:2112.09118},
  year={2021}
}

@inproceedings{gao2023precise,
  title={Precise zero-shot dense retrieval without relevance labels},
  author={Gao, Luyu and Ma, Xueguang and Lin, Jimmy and Callan, Jamie},
  booktitle={Proceedings of the 61st Annual Meeting of the Association for Computational Linguistics (Volume 1: Long Papers)},
  pages={1762--1777},
  year={2023}
}

@inproceedings{weller2024nevir,
  title={Nevir: Negation in neural information retrieval},
  author={Weller, Orion and Lawrie, Dawn and Van Durme, Benjamin},
  booktitle={Proceedings of the 18th Conference of the European Chapter of the Association for Computational Linguistics (Volume 1: Long Papers)},
  pages={2274--2287},
  year={2024}
}

@inproceedings{zhang2025excluir,
  title={Excluir: Exclusionary neural information retrieval},
  author={Zhang, Wenhao and Zhang, Mengqi and Wu, Shiguang and Pei, Jiahuan and Ren, Zhaochun and De Rijke, Maarten and Chen, Zhumin and Ren, Pengjie},
  booktitle={Proceedings of the AAAI Conference on Artificial Intelligence},
  volume={39},
  number={12},
  pages={13295--13303},
  year={2025}
}

@inproceedings{xiao2024c,
  title={C-pack: Packed resources for general chinese embeddings},
  author={Xiao, Shitao and Liu, Zheng and Zhang, Peitian and Muennighoff, Niklas and Lian, Defu and Nie, Jian-Yun},
  booktitle={Proceedings of the 47th international ACM SIGIR conference on research and development in information retrieval},
  pages={641--649},
  year={2024}
}

@article{xiong2020approximate,
  title={Approximate nearest neighbor negative contrastive learning for dense text retrieval},
  author={Xiong, Lee and Xiong, Chenyan and Li, Ye and Tang, Kwok-Fung and Liu, Jialin and Bennett, Paul and Ahmed, Junaid and Overwijk, Arnold},
  journal={arXiv preprint arXiv:2007.00808},
  year={2020}
}

@inproceedings{qu2021rocketqa,
  title={RocketQA: An optimized training approach to dense passage retrieval for open-domain question answering},
  author={Qu, Yingqi and Ding, Yuchen and Liu, Jing and Liu, Kai and Ren, Ruiyang and Zhao, Wayne Xin and Dong, Daxiang and Wu, Hua and Wang, Haifeng},
  booktitle={Proceedings of the 2021 conference of the North American chapter of the association for computational linguistics: human language technologies},
  pages={5835--5847},
  year={2021}
}

@inproceedings{reimers2019sentence,
  title={Sentence-bert: Sentence embeddings using siamese bert-networks},
  author={Reimers, Nils and Gurevych, Iryna},
  booktitle={Proceedings of the 2019 conference on empirical methods in natural language processing and the 9th international joint conference on natural language processing (EMNLP-IJCNLP)},
  pages={3982--3992},
  year={2019}
}

@article{oord2018representation,
  title={Representation learning with contrastive predictive coding},
  author={Oord, Aaron van den and Li, Yazhe and Vinyals, Oriol},
  journal={arXiv preprint arXiv:1807.03748},
  year={2018}
}

@article{thakur2021beir,
  title={Beir: A heterogenous benchmark for zero-shot evaluation of information retrieval models},
  author={Thakur, Nandan and Reimers, Nils and R{\"u}ckl{\'e}, Andreas and Srivastava, Abhishek and Gurevych, Iryna},
  journal={arXiv preprint arXiv:2104.08663},
  year={2021}
}

@book{robertson2009probabilistic,
  title={The probabilistic relevance framework: BM25 and beyond},
  author={Robertson, Stephen and Zaragoza, Hugo},
  volume={4},
  year={2009},
  publisher={Now Publishers Inc}
}

@article{miller1995wordnet,
  title={WordNet: a lexical database for English},
  author={Miller, George A},
  journal={Communications of the ACM},
  volume={38},
  number={11},
  pages={39--41},
  year={1995},
  publisher={ACM New York, NY, USA}
}

@article{lewis2020retrieval,
  title={Retrieval-augmented generation for knowledge-intensive nlp tasks},
  author={Lewis, Patrick and Perez, Ethan and Piktus, Aleksandra and Petroni, Fabio and Karpukhin, Vladimir and Goyal, Naman and K{\"u}ttler, Heinrich and Lewis, Mike and Yih, Wen-tau and Rockt{\"a}schel, Tim and others},
  journal={Advances in neural information processing systems},
  volume={33},
  pages={9459--9474},
  year={2020}
}

@article{liu2024lost,
  title={Lost in the middle: How language models use long contexts},
  author={Liu, Nelson F and Lin, Kevin and Hewitt, John and Paranjape, Ashwin and Bevilacqua, Michele and Petroni, Fabio and Liang, Percy},
  journal={Transactions of the association for computational linguistics},
  volume={12},
  pages={157--173},
  year={2024}
}

\appendix

\section{Additional Dataset Construction Details}
\label{app:dataset_construction}

We construct three diagnostic test sets to evaluate whether a retriever can distinguish constraint-satisfying evidence from topically similar constraint-violating evidence. The Antonym set is built from SciFact, the Negation set from SciDocs, and the Exclusion set from NFCorpus. Table~\ref{tab:diagnostic_dataset_stats} summarizes the resulting dataset sizes.

\begin{table}[h]
    \centering
    \small
    \begin{tabular}{lccc}
        \toprule
        Dataset & Source & Queries & Documents \\
        \midrule
        Antonym & SciFact & 102 & 306 \\
        Negation & SciDocs & 104 & 304 \\
        Exclusion & NFCorpus & 104 & 312 \\
        \bottomrule
    \end{tabular}
    \caption{Statistics of the self-constructed diagnostic datasets.}
    \label{tab:diagnostic_dataset_stats}
\end{table}

For each source corpus, we first select candidate documents with sufficient length and segment them into shorter sub-documents. For the Antonym set, we identify a keyword with a valid antonym and construct paired queries and paired documents by replacing the keyword with its antonym while keeping the surrounding topic fixed. For the Negation set, we construct paired examples by changing whether a claim or relation is asserted or negated. For the Exclusion set, we construct queries that require excluding a specific attribute, entity, or condition, together with documents that either satisfy or violate the exclusion.

Each query is assigned three types of evidence labels. A constraint-satisfying document matches the query topic and satisfies the stated constraint. A constraint-violating document matches the topic but supports the opposite constraint direction. A neutral topical document is related to the query topic but does not explicitly satisfy or violate the constraint. For nDCG computation, we assign graded relevance scores of 2 to satisfying documents, 1 to neutral topical documents, and 0 to violating documents. For violation-oriented metrics, only constraint-violating documents are counted as violations.
\section{Constraint Compatibility Encoder Training Details}
\label{app:constraint_encoder_training}

The constraint compatibility encoder $E_C$ is initialized from \texttt{bge-large-en-v1.5} and trained as a bi-encoder with mean pooling. We set the maximum sequence length to 128. We use a two-stage training recipe. In the first stage, we train on WordNet-derived word-level polarity triplets, where the positive and negative examples differ by antonymy or polarity direction. This stage provides a basic lexical polarity signal before sentence-level training~\cite{miller1995wordnet}.

In the second stage, we continue training on sentence-level triplets constructed from NevIR and ExcluIR. We sample 800 training queries from NevIR and 800 training queries from ExcluIR. Each training instance is a triplet $(q,d^+,d^-)$, where $d^+$ satisfies the query constraint and $d^-$ is topically related but constraint-violating. These sentence-level triplets teach $E_C$ to score compatibility at the passage level rather than only at the word level.

To avoid data leakage, all queries and triples used to train $E_C$ are removed from the corresponding evaluation splits. They do not appear in any reported NevIR or ExcluIR test results. We use the same BGE-style query prefix during training and inference. 
The NevIR and ExcluIR results are held-out in-domain evaluations rather than fully out-of-domain transfer.
The model is optimized with the multiple negatives ranking loss described in Section~\ref{subsec:lexical_polarity_supervision}, using the explicit violating document and in-batch negatives as contrastive negatives. We train for one epoch with a batch size of 16, using AdamW with a learning rate of $2\times 10^{-5}$ throughout both stages.

\section{NegConstraint Experiment Result}
\label{app:negconstraint}

\begin{table}[h]
    \small
    \centering
    \resizebox{\columnwidth}{!}{
    \begin{tabular}{l|ccc}
        \toprule
        \multicolumn{1}{c|}{Methods}
        & Recall@5$\uparrow$
        & nDCG@5$\uparrow$
        & MAP@10$\uparrow$ \\
        \midrule
        NS-IR      & 95.96 & 76.64 & 70.05 \\
        \midrule
        CoDeR-Seq & 96.46 & 81.20 & 76.00 \\
        CoDeR-Union & \textbf{96.97} & \textbf{81.61} & \textbf{76.38} \\
        \bottomrule
    \end{tabular}
    }
    \caption{
Results on the released NegConstraint benchmark.
Following the released evaluation protocol, we report Recall@5, nDCG@5, and MAP@10.
Higher values are better.
}
    \label{tab:negconstraint}
\end{table}

\section{Full Efficiency Accounting}
\label{app:full_efficiency_accounting}

Table~\ref{tab:full_efficiency_accounting} reports the per-setting efficiency measurements used in Section~\ref{subsec:efficiency_analysis}. 
Time is reported in seconds, API cost in US dollars, and token usage in thousands.

\begin{table}[t]
    \centering
    \scriptsize
    \setlength{\tabcolsep}{4pt}
    \begin{tabular}{llrrr}
        \toprule
        Setting & Method & Time(s)$\downarrow$ & Cost(\$)$\downarrow$ & Tokens(K)$\downarrow$ \\
        \midrule
         & HyDE & 3900.98 & 1.22 & 146.3 \\
         & NS-IR & 7190.02 & 0.85 & 128.8 \\
        Antonym & DEO & 180.94 & 0.18 & 50.5 \\
         & CoDeR-Seq & 6.14 & 0.00 & 0.0 \\
         & CoDeR-Union & 6.19 & 0.00 & 0.0 \\
        \midrule
         & HyDE & 3222.89 & 1.24 & 148.8 \\
         & NS-IR & 6979.14 & 0.98 & 144.0 \\
        Negation & DEO & 187.12 & 0.18 & 50.9 \\
         & CoDeR-Seq & 6.44 & 0.00 & 0.0 \\
         & CoDeR-Union & 6.81 & 0.00 & 0.0 \\
        \midrule
         & HyDE & 4232.36 & 1.00 & 124.1 \\
         & NS-IR & 8249.47 & 1.27 & 177.9 \\
        Exclusion & DEO & 190.24 & 0.19 & 51.9 \\
         & CoDeR-Seq & 7.36 & 0.00 & 0.0 \\
         & CoDeR-Union & 7.72 & 0.00 & 0.0 \\
        \midrule
         & HyDE & 56204.29 & 9.81 & 1183.0 \\
         & NS-IR & 61621.02 & 5.18 & 755.3 \\
        NevIR & DEO & 16516.02 & 1.70 & 487.7 \\
         & CoDeR-Seq & 131.23 & 0.00 & 0.0 \\
         & CoDeR-Union & 65.95 & 0.00 & 0.0 \\
        \midrule
         & HyDE & 69233.49 & 13.85 & 1668.9 \\
         & NS-IR & 69775.07 & 10.65 & 1501.5 \\
        ExcluIR & DEO & 17620.81 & 2.03 & 531.4 \\
         & CoDeR-Seq & 66.70 & 0.00 & 0.0 \\
         & CoDeR-Union & 89.09 & 0.00 & 0.0 \\
        \midrule
         & HyDE & 31213.27 & 5.18 & 582.9 \\
         & NS-IR & 39738.51 & 13.28 & 1823.8 \\
        NegConstraint & DEO & 4782.49 & 0.83 & 204.4 \\
         & CoDeR-Seq & 82.37 & 0.00 & 0.0 \\
         & CoDeR-Union & 86.74 & 0.00 & 0.0 \\
        \midrule
         & HyDE & 716.90 & 0.31 & 35.8 \\
         & NS-IR & 732.22 & 0.13 & 19.8 \\
        End-to-End & DEO & 77.73 & 0.04 & 10.4 \\
         & CoDeR-Seq & 1.58 & 0.00 & 0.0 \\
         & CoDeR-Union & 0.94 & 0.00 & 0.0 \\
        \bottomrule
    \end{tabular}
    \caption{
    Full per-setting efficiency accounting.
    Time is reported in seconds.
    Cost is estimated API cost in US dollars.
    Tokens are API tokens consumed during inference, reported in thousands.
    The End-to-End setting is included for completeness but is not part of the aggregate efficiency table in Section~\ref{subsec:efficiency_analysis}.
    }
    \label{tab:full_efficiency_accounting}
\end{table}
\section{Violation Survival Analysis}
\label{app:violation_survival}

We provide rank-wise violation survival curves as a complementary view of the diagnostic results in Table~\ref{tab:selfbuilt}.
For each query, the first violating rank records the earliest retrieved position at which a constraint-violating document appears.
The survival value at rank $k$ is the fraction of queries whose first violating rank is greater than $k$, i.e., the query has not exposed violating evidence within the first $k$ retrieved documents.
Higher curves therefore indicate safer early retrieved contexts.

\begin{figure*}[t]
    \centering
    \includegraphics[width=0.95\textwidth]{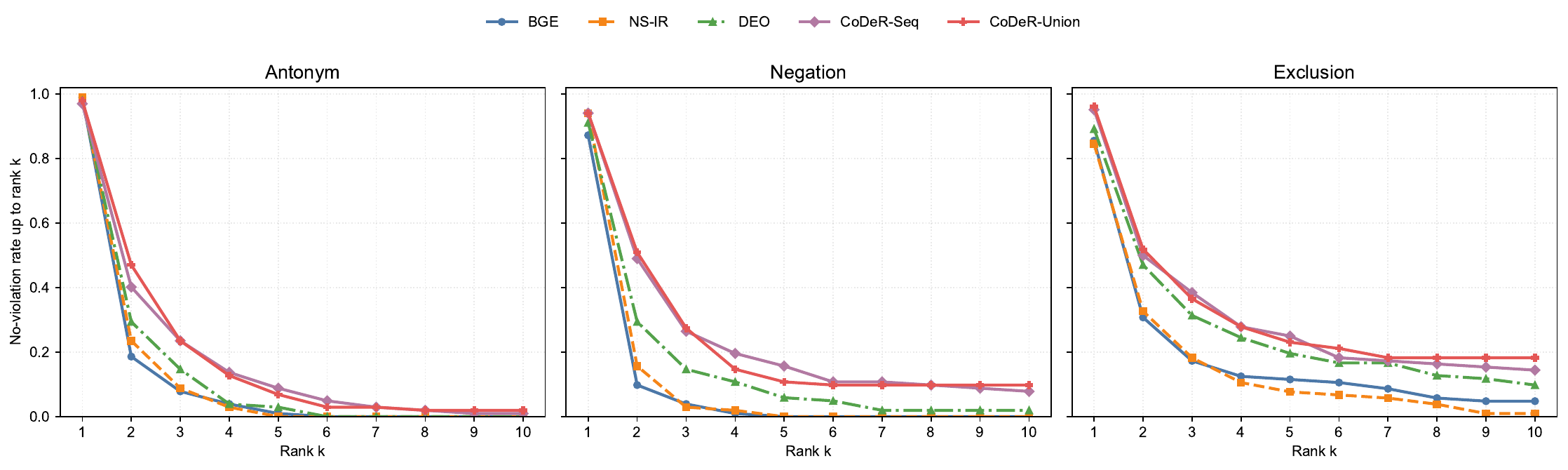}
    \caption{
    Violation survival curves on the diagnostic Antonym, Negation, and Exclusion datasets.
    The y-axis reports the fraction of queries with no constraint-violating evidence up to rank $k$.
    Higher curves indicate that violations are delayed deeper in the ranked list.
    CoDeR variants keep higher no-violation rates at early ranks, complementing the V@$k$ and FVR results in Table~\ref{tab:selfbuilt}.
    }
    \label{fig:violation_survival_curve}
\end{figure*}

Figure~\ref{fig:violation_survival_curve} shows that CoDeR-Seq and CoDeR-Union generally maintain higher no-violation rates at early ranks than topical or query-rewriting baselines.
This indicates that CoDeR changes the timing of violation exposure rather than only improving an aggregate FVR value.
By delaying the first constraint-violating document, compatibility-aware retrieval reduces the chance that downstream systems consuming only the top few retrieved documents encounter directionally wrong evidence.

\section{Hyperparameter Policy Sensitivity}
\label{app:policy_sensitivity}

\begin{figure}[t]
    \centering
    \includegraphics[width=\columnwidth]{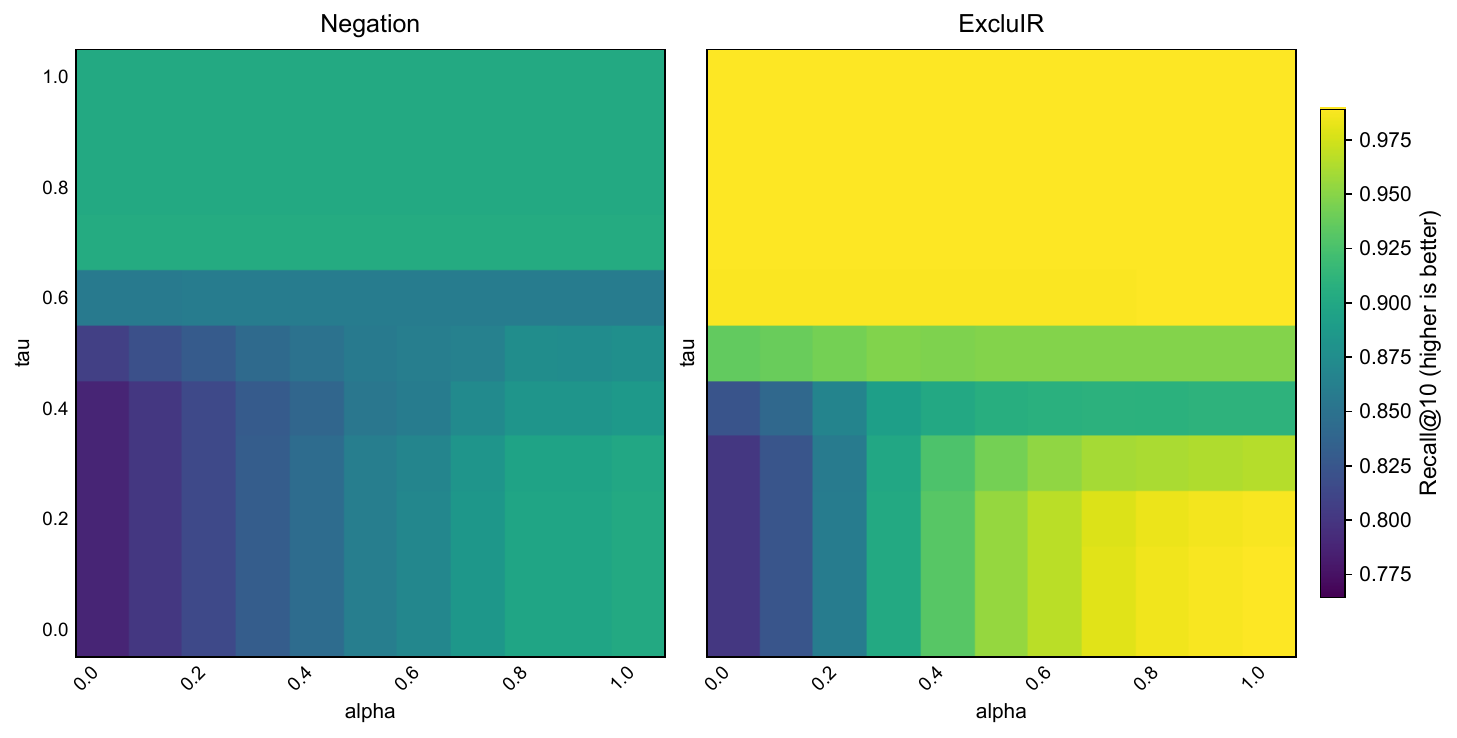}
    \caption{CoDeR-Seq policy sensitivity measured by Recall@10 on Negation and ExcluIR. Higher values indicate better preservation of satisfying evidence. The map shows that recall remains high over a broad low-to-moderate threshold region, but can drop when thresholding becomes too restrictive.}
    \label{fig:policy_seq_recall}
\end{figure}

\begin{figure}[t]
    \centering
    \includegraphics[width=\columnwidth]{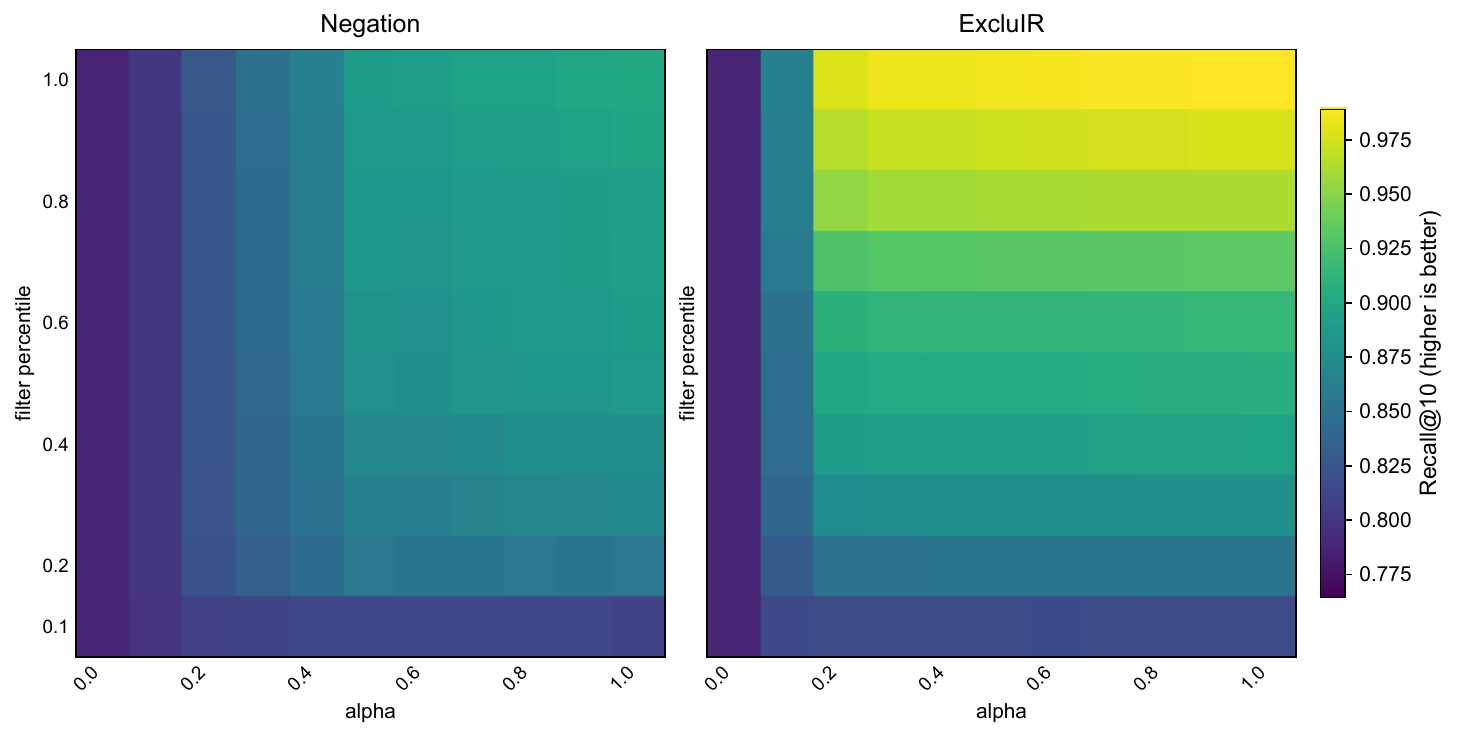}
    \caption{CoDeR-Union policy sensitivity measured by Recall@10 on Negation and ExcluIR. The relative filter percentile creates smoother high-recall regions, especially when $\alpha$ keeps sufficient topical-retrieval weight.}
    \label{fig:policy_union_recall}
\end{figure}

\begin{figure}[t]
    \centering
    \includegraphics[width=\columnwidth]{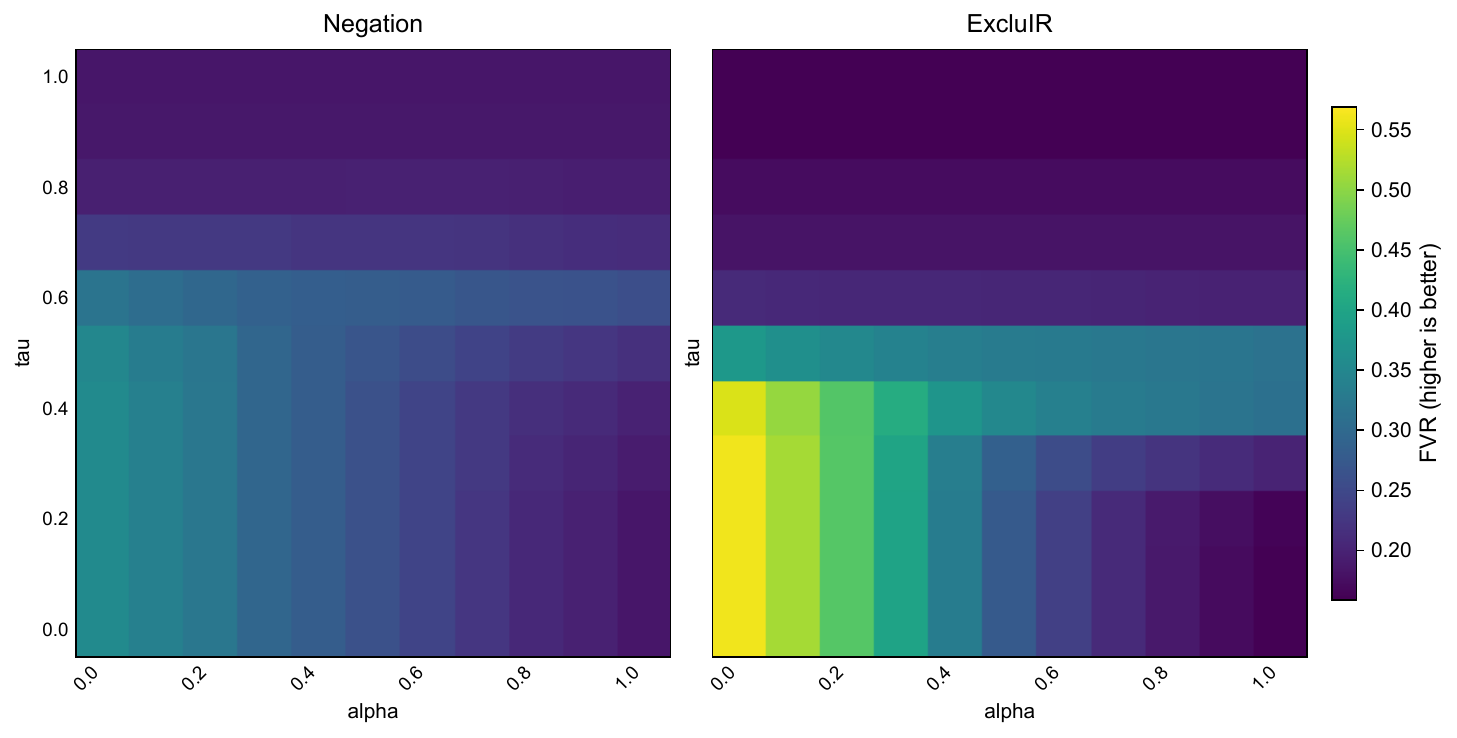}
    \caption{CoDeR-Seq policy sensitivity measured by FVR on Negation and ExcluIR. Each cell shows FVR normalized by $k_{\max}+1=11$; values closer to 1 indicate that the first violation appears later in the ranked list, and 1 indicates no violation within the top 10. The absolute threshold $\tau$ can improve violation delay in effective regions, but its behavior depends on dataset-specific compatibility-score calibration.}
    \label{fig:policy_seq_fvr}
\end{figure}

\begin{figure}[t]
    \centering
    \includegraphics[width=\columnwidth]{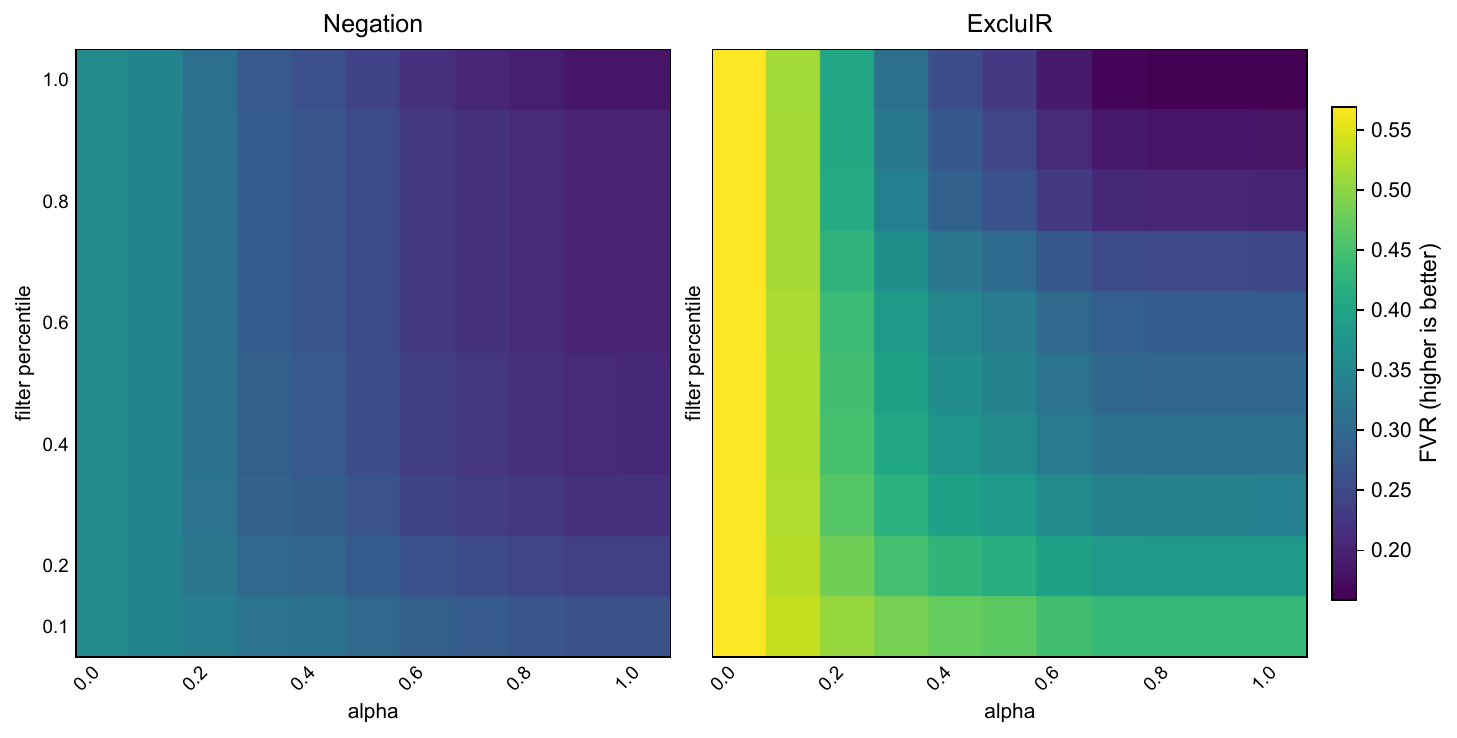}
    \caption{CoDeR-Union policy sensitivity measured by FVR on Negation and ExcluIR. Each cell shows FVR normalized by $k_{\max}+1=11$; values closer to 1 indicate that the first violation appears later in the ranked list, and 1 indicates no violation within the top 10. The relative filter percentile produces gradual changes in violation delay: stronger filtering tends to push violations later, while weaker filtering preserves a larger candidate set.}
    \label{fig:policy_union_fvr}
\end{figure}

We provide the full policy-sensitivity heatmaps for CoDeR-Seq and CoDeR-Union.
Figures~\ref{fig:policy_seq_recall} and~\ref{fig:policy_union_recall} report the Recall@10 sensitivity patterns for CoDeR-Seq and CoDeR-Union, respectively, while Figures~\ref{fig:policy_seq_fvr} and~\ref{fig:policy_union_fvr} report the corresponding FVR sensitivity patterns.
The analysis uses Negation and ExcluIR as representative datasets and reports two complementary metrics.
Recall@10 measures whether satisfying evidence remains in the retrieved list, while FVR measures how far the first constraint-violating document is pushed down the ranking.
For the FVR heatmaps, each cell reports the raw FVR normalized by $k_{\max}+1$ with $k_{\max}=10$, so a value of 1 indicates that no violation is found within the top 10.
For CoDeR-Seq, the grid varies the interpolation weight $\alpha$ and the absolute compatibility threshold $\tau$.
For CoDeR-Union, the grid varies $\alpha$ and the relative filter percentile.
In the six BEIR-style datasets, CoDeR-Seq uses $\alpha=0.5$ and $\tau=0.9$, while CoDeR-Union uses $\alpha=0.8$ and a relative filter percentile of 0.8.
For the other datasets, CoDeR-Seq uses $\alpha=0.2$ and $\tau=0.3$, while CoDeR-Union uses $\alpha=0.3$ and a relative filter percentile of 0.1.

Together, Figures~\ref{fig:policy_seq_recall}--\ref{fig:policy_union_fvr} show that CoDeR is not driven by a single accidental setting.
CoDeR-Seq has effective bands over $\alpha$ and $\tau$, but because $\tau$ is an absolute compatibility threshold, the useful region can shift with score calibration across datasets.
CoDeR-Union replaces this absolute decision with relative filtering, making the policy less dependent on raw score scale and producing smoother behavior across Negation and ExcluIR.
The endpoint behavior of $\alpha$ is also consistent with the intended mechanism: larger topical weight preserves recall, while stronger compatibility-driven filtering delays early violation exposure.

\section{Ablation Study}
\label{app:ablation_study}

We further conduct ablation experiments to test the structural claim behind CoDeR rather than only to compare component strength. The main paper argues that constraint-sensitive retrieval fails when topical relevance and constraint compatibility are collapsed into a single semantic-similarity score: a violating document can be highly topical because it mentions the same entities, attributes, and domain vocabulary, while still supporting the wrong constraint direction. The ablations therefore ask whether CoDeR's gains come from the proposed decoupling of topicality and compatibility, or from simpler alternatives such as using a stronger topical retriever, replacing the compatibility side with another dense encoder, relying on the trained constraint encoder alone, or attaching a standard reranker after retrieval.

Tables~\ref{tab:ablation_v1} and~\ref{tab:ablation_v2} examine the two encoder roles in this decomposition. In Group~1, we fix the compatibility-side encoder as CoDeR and vary the topical encoder among BGE-large, BGE-base, and BGE-small. This setting tests whether the compatibility signal remains meaningful when the candidate generator changes. In Group~2, we fix the topical encoder as BGE-large and replace the compatibility-side backbone with Contriever or miniLM. This setting tests whether a generic semantic encoder can play the same role as a constraint-compatibility encoder trained with satisfying--violating contrasts.

The first insight is that Encoder~A behaves like a coverage component, not the source of constraint direction. Stronger topical encoders generally preserve ordinary BEIR-style retrieval better on SciFact and NFCorpus, which is expected because they determine which topically plausible candidates enter the pool. However, when Encoder~B is kept as the trained compatibility encoder, changing Encoder~A produces much smaller shifts in early-violation behavior than changing Encoder~B. This is important because it rules out the interpretation that CoDeR works simply because BGE-large is a strong retriever. A strong topical encoder can supply better candidate coverage, but it does not by itself decide whether a topically similar document satisfies or violates the user's constraint.

The second insight is that Encoder~B is not just another semantic reranker. When the compatibility side is replaced by Contriever or miniLM, the system may still rank semantically related documents, but it loses the specific pressure that pushes wrong-direction evidence downward. This exposes the difference between semantic relatedness and compatibility: Contriever and miniLM can recognize that a document is about the same topic, yet they are not trained to treat satisfying and violating counterparts as opposite retrieval outcomes. The occasional improvement of miniLM on a narrow setting should therefore not be read as evidence that any dense encoder can substitute for $E_C$; rather, it shows that aggressive reordering can sometimes remove violations while also weakening the intended topical-preservation behavior.

Together, Tables~\ref{tab:ablation_v1} and~\ref{tab:ablation_v2} support the central design choice of CoDeR. The modular split is useful because the two encoders answer different questions: Encoder~A asks whether a document is about the query, while Encoder~B asks whether the document is compatible with the query's constraint direction. The observed robustness across topical backbones and the degradation under generic compatibility replacements jointly indicate that constraint-aware retrieval is not obtained by merely scaling topical retrieval. It requires a learned compatibility dimension that is composed with topical coverage at retrieval time.

\subsection{Compatibility Encoder Score Separation}

\begin{figure}[t]
    \centering
    \includegraphics[width=\columnwidth]{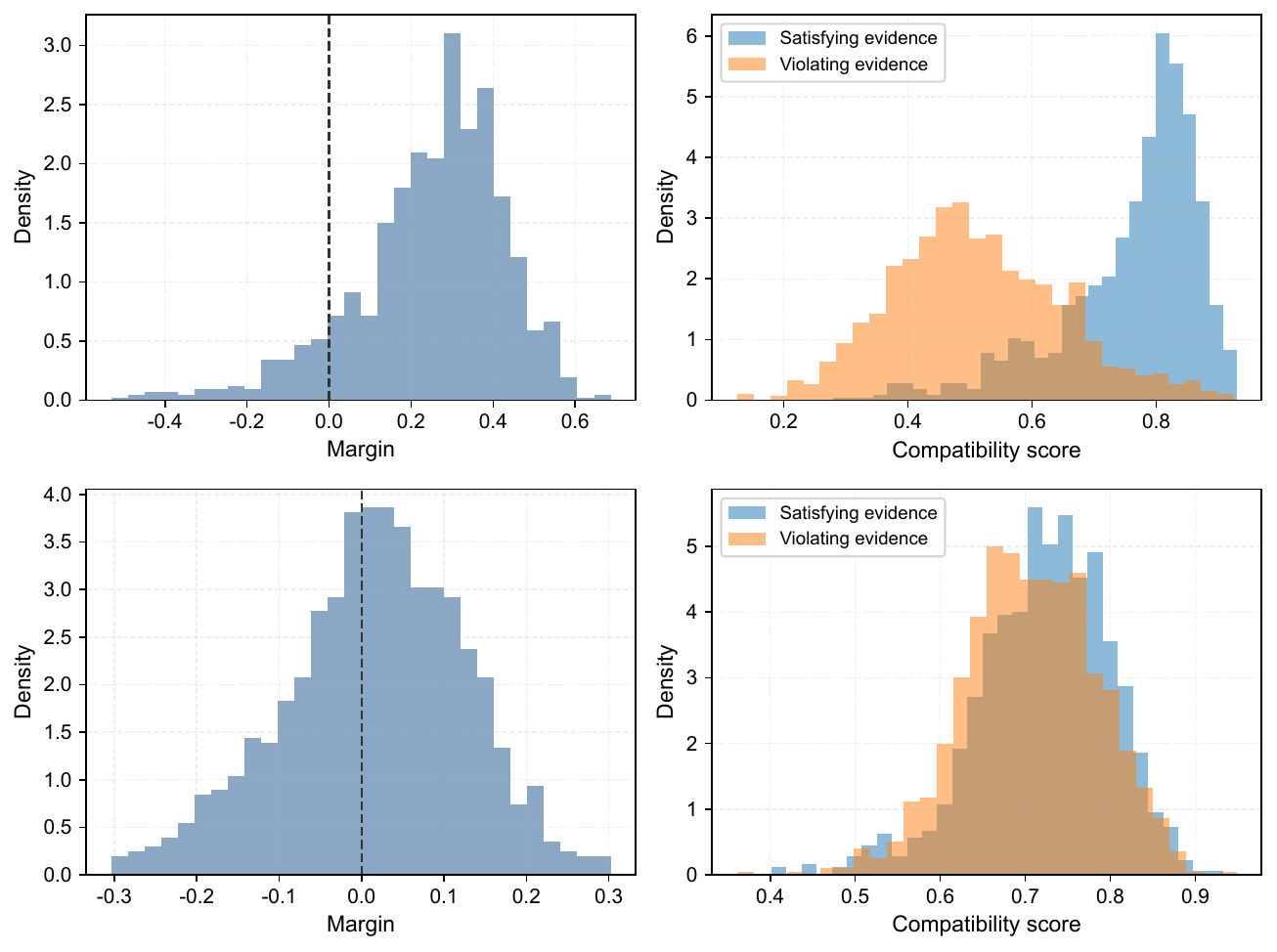}
    \caption{
    Compatibility-score separation on ExcluIR.
    The top row uses the trained BGE constraint encoder, and the bottom row uses the base BGE-large encoder.
    Left panels show query-level satisfying-minus-violating margins; right panels show raw compatibility-score distributions for satisfying and violating evidence.
    The trained compatibility encoder produces more positive query-level margins and clearer score separation than the base topical encoder.
    }
    \label{fig:encoder_b_score_separation}
\end{figure}

We further inspect whether the trained compatibility encoder changes the score geometry relative to the base BGE encoder.
Figure~\ref{fig:encoder_b_score_separation} compares the trained BGE constraint encoder with BGE-large on ExcluIR using query-level satisfying-minus-violating margins and raw compatibility-score distributions.
The trained encoder shifts the margin distribution toward positive values and assigns higher compatibility scores to satisfying evidence than to violating evidence.
By contrast, the base BGE encoder shows more overlap between the two evidence types, consistent with its role as a topical semantic encoder rather than a constraint-compatibility scorer.
This analysis complements the component ablations below by showing that the trained Encoder~B contributes a separable compatibility direction in the scoring space.

\begin{table*}[h]
    \small
    \centering
    \resizebox{\textwidth}{!}{
    \begin{tabular}{ll|cc|cc|ccccc|ccccc|ccccc}
        \toprule
        \multirow{2.5}{*}{Encoder A} & \multirow{2.5}{*}{Encoder B}
        & \multicolumn{2}{c|}{SciFact} & \multicolumn{2}{c|}{NFCorpus}
        & \multicolumn{5}{c|}{Antonym} & \multicolumn{5}{c|}{Negation} & \multicolumn{5}{c}{Exclusion} \\
        \cmidrule{3-4}\cmidrule{5-6}\cmidrule{7-11}\cmidrule{12-16}\cmidrule{17-21}
        & & \makecell{nDCG}$\uparrow$ & \makecell{MAP}$\uparrow$
          & \makecell{nDCG}$\uparrow$ & \makecell{MAP}$\uparrow$
          & V@2$\downarrow$ & V@3$\downarrow$ & V@5$\downarrow$ & V@10$\downarrow$ & FVR$\uparrow$
          & V@2$\downarrow$ & V@3$\downarrow$ & V@5$\downarrow$ & V@10$\downarrow$ & FVR$\uparrow$
          & V@2$\downarrow$ & V@3$\downarrow$ & V@5$\downarrow$ & V@10$\downarrow$ & FVR$\uparrow$ \\
        \midrule
        \multicolumn{21}{l}{\textit{Group 1: Fix Encoder B = CoDeR-Seq, vary Encoder A}} \\
        \midrule
        BGE-large  & CoDeR-Seq & \textbf{74.77} & \textbf{69.93} & \textbf{37.24} & \textbf{28.00} & 59.80 & 76.47 & 91.18 & \textbf{99.02} & 2.95 & \textbf{50.98} & \textbf{73.53} & \textbf{84.31} & 92.16 & \textbf{3.53} & 50.00 & 60.58 & \textbf{74.04} & \textbf{84.62} & \textbf{4.26} \\
        BGE-base   & CoDeR-Seq & 73.88 & 68.88 & 36.68 & 27.54 & 60.78 & \textbf{74.51} & 93.14 & \textbf{99.02} & 2.94 & 51.96 & 74.51 & \textbf{84.31} & 94.12 & 3.42 & 50.98 & 64.71 & 77.45 & 86.27 & 4.05 \\
        BGE-small  & CoDeR-Seq & 69.68 & 64.92 & 34.82 & 25.57 & \textbf{58.82} & 77.45 & \textbf{90.20} & \textbf{99.02} & \textbf{2.99} & 51.96 & 74.51 & \textbf{84.31} & 92.16 & 3.50 & 49.02 & 63.73 & 75.49 & 86.27 & 4.14 \\
        \midrule
        \multicolumn{21}{l}{\textit{Group 2: Fix Encoder A = BGE-large, vary Encoder B}} \\
        \midrule
        BGE-large  & Contriever          & \textbf{74.77} & \textbf{69.93} & \textbf{37.24} & \textbf{28.00} & 80.39 & 94.12 & 99.02 & 100.00 & 2.26 & 71.57 & 83.33 & 95.10 & 98.04 & 2.66 & 66.67 & 82.35 & 92.16 & 97.06 & 2.77 \\
        BGE-large  & miniLM              & \textbf{74.77} & \textbf{69.93} & \textbf{37.24} & \textbf{28.00} & 76.47 & 86.27 & 92.16 & \textbf{99.02} & 2.64 & 59.80 & \textbf{73.53} & \textbf{84.31} & \textbf{89.22} & 3.51 & \textbf{45.20} & \textbf{56.86} & 79.41 & 88.24 & 4.09 \\
        \bottomrule
    \end{tabular}
    }
    \caption{Ablation study for CoDeR-Seq. nDCG and MAP are reported at @10. Group~1 fixes Encoder~B as CoDeR-Seq and varies Encoder~A among BGE-large, BGE-base, and BGE-small. Group~2 fixes Encoder~A as BGE-large and varies Encoder~B among Contriever and miniLM. The BGE-large / CoDeR-Seq row at the end of Group~1 serves as the shared baseline for both groups.}
    \label{tab:ablation_v1}
\end{table*}

\begin{table*}[h]
    \small
    \centering
    \resizebox{\textwidth}{!}{
    \begin{tabular}{ll|cc|cc|ccccc|ccccc|ccccc}
        \toprule
        \multirow{2.5}{*}{Encoder A} & \multirow{2.5}{*}{Encoder B}
        & \multicolumn{2}{c|}{SciFact} & \multicolumn{2}{c|}{NFCorpus}
        & \multicolumn{5}{c|}{Antonym} & \multicolumn{5}{c|}{Negation} & \multicolumn{5}{c}{Exclusion} \\
        \cmidrule{3-4}\cmidrule{5-6}\cmidrule{7-11}\cmidrule{12-16}\cmidrule{17-21}
        & & \makecell{nDCG}$\uparrow$ & \makecell{MAP}$\uparrow$
          & \makecell{nDCG}$\uparrow$ & \makecell{MAP}$\uparrow$
          & V@2$\downarrow$ & V@3$\downarrow$ & V@5$\downarrow$ & V@10$\downarrow$ & FVR$\uparrow$
          & V@2$\downarrow$ & V@3$\downarrow$ & V@5$\downarrow$ & V@10$\downarrow$ & FVR$\uparrow$
          & V@2$\downarrow$ & V@3$\downarrow$ & V@5$\downarrow$ & V@10$\downarrow$ & FVR$\uparrow$ \\
        \midrule
        \multicolumn{21}{l}{\textit{Group 1: Fix Encoder B = CoDeR-Union, vary Encoder A}} \\
        \midrule
        BGE-large  & CoDeR-Union & \textbf{74.91} & \textbf{70.17} & 36.36 & 27.12 & \textbf{52.94} & 76.47 & \textbf{93.13} & 98.04 & \textbf{3.00} & 49.02 & 72.55 & 89.22 & 90.20 & 3.47 & 48.08 & 61.54 & 75.96 & 81.73 & 4.37 \\
        BGE-base   & CoDeR-Union & 74.05 & 69.06 & \textbf{36.59} & \textbf{27.40} & 53.92 & \textbf{75.49} & 96.08 & 98.04 & 2.93 & \textbf{46.08} & 72.55 & 90.20 & 90.20 & 3.50 & 49.02 & 65.69 & 79.41 & 84.31 & 4.15 \\
        BGE-small  & CoDeR-Union & 69.92 & 65.28 & 35.31 & 25.86 & \textbf{52.94} & 76.47 & 94.12 & 98.04 & 2.98 & 47.06 & 72.55 & 89.22 & 90.20 & 3.49 & 49.02 & 62.75 & 79.41 & 84.31 & 4.17 \\
        \midrule
        \multicolumn{21}{l}{\textit{Group 2: Fix Encoder A = BGE-large, vary Encoder B}} \\
        \midrule
        BGE-large  & Contriever          & 68.27 & 64.36 & 32.86 & 23.34 & 74.51 & 90.20 & 99.02 & 100.00 & 2.19 & 46.47 & 80.39 & \textbf{85.10} & 89.22 & 3.00 & 55.88 & 79.41 & 90.20 & 94.12 & 3.10 \\
        BGE-large  & miniLM              & 71.38 & 66.96 & 33.44 & 23.64 & 66.67 & 80.39 & 94.12 & \textbf{97.06} & 2.81 & 51.96 & \textbf{69.61} & 87.25 & \textbf{88.24} & \textbf{3.63} & \textbf{33.33} & \textbf{47.06} & \textbf{70.59} & \textbf{77.45} & \textbf{4.99} \\
        \bottomrule
    \end{tabular}
    }
    \caption{Ablation study for CoDeR-Union. nDCG and MAP are reported at @10. Group~1 fixes Encoder~B as CoDeR-Union and varies Encoder~A among BGE-large, BGE-base, and BGE-small. Group~2 fixes Encoder~A as BGE-large and varies Encoder~B among Contriever and miniLM. The BGE-large / CoDeR-Union row at the end of Group~1 serves as the shared baseline for both groups.}
    \label{tab:ablation_v2}
\end{table*}

\begin{table}[h]
    \small
    \centering
    \resizebox{\columnwidth}{!}{
    \begin{tabular}{l|ccc}
        \toprule
        \multirow{2}{*}{Model} & \multicolumn{3}{c}{NevIR} \\
        \cmidrule{2-4}
        & Recall@5$\uparrow$ & V@1$\downarrow$ & FVR$\uparrow$ \\
        \midrule
        BGE-large & \textbf{85.85} & 19.6 & 3.70 \\
        Trained BGE Constraint Encoder & 79.2 & 14.2 & 4.45 \\
        CoDeR-Seq & 81.65 & 15.4 & 4.26 \\
        CoDeR-Union & 81.7 & \textbf{13.9} & \textbf{4.28} \\
        \bottomrule
    \end{tabular}
    }
    \caption{Comparison of BGE-large, the trained BGE constraint encoder, CoDeR-Seq, and CoDeR-Union on NevIR.}
    \label{tab:ablation_nevir_encoder_comparison}
\end{table}

Table~\ref{tab:ablation_nevir_encoder_comparison} isolates the role of the trained compatibility encoder on NevIR, where the relevant failure is not topic mismatch but opposite-direction negation evidence appearing too early. This ablation separates three possibilities that are conflated in the full system: a strong topical encoder alone, the trained constraint encoder alone, and the integrated CoDeR policies. BGE-large represents the first case: it preserves topical recall because it is optimized for semantic matching, but this objective does not directly encode which member of a negation pair satisfies the query. The trained BGE constraint encoder represents the second case: it shows that polarity supervision can reshape the score space toward compatibility, but using it alone removes part of the topical-retrieval function that a general retriever still needs.

The CoDeR rows are therefore the critical comparison rather than a simple middle point between BGE-large and the constraint encoder. CoDeR-Seq and CoDeR-Union use the compatibility signal as a retrieval-side control over a topical candidate process, so the learned constraint direction is not treated as a replacement for topicality but as an additional axis for ranking and filtering. This explains the intended trade-off in Table~\ref{tab:ablation_nevir_encoder_comparison}: the full systems give up some pure topical recall relative to BGE-large, but they reduce early violation exposure while retaining substantially more retrieval coverage than a compatibility-only interpretation would require. The insight is that $E_C$ is useful because it becomes part of a two-signal retrieval policy, not because a standalone constraint encoder should replace the topical retriever.

\begin{table*}[h]
    \scriptsize
    \centering
    \resizebox{\textwidth}{!}{
    \begin{tabular}{l|cc|cc|cc|cc|cc}
        \toprule
        \multirow{2}{*}{Model}
        & \multicolumn{2}{c|}{Antonym}
        & \multicolumn{2}{c|}{Negation}
        & \multicolumn{2}{c|}{Exclusion}
        & \multicolumn{2}{c|}{NevIR}
        & \multicolumn{2}{c}{ExcluIR} \\
        \cmidrule{2-3}\cmidrule{4-5}\cmidrule{6-7}\cmidrule{8-9}\cmidrule{10-11}
        & nDCG@5$\uparrow$ & FVR$\uparrow$
        & nDCG@5$\uparrow$ & FVR$\uparrow$
        & nDCG@5$\uparrow$ & FVR$\uparrow$
        & nDCG@5$\uparrow$ & FVR$\uparrow$
        & nDCG@5$\uparrow$ & FVR$\uparrow$ \\
        \midrule
        BGE-large+BGE Reranker
        & 80.00 & 4.32
        & \textbf{79.57} & 2.90
        & \textbf{75.34} & 2.91
        & 60.48 & 4.00
        & 83.87 & 1.79 \\

        CoDeR-Seq+BGE Reranker
        & \textbf{80.43} & 4.42
        & 78.83 & 3.42
        & 75.07 & 3.63
        & \textbf{60.78} & \textbf{4.28}
        & 85.35 & 4.11 \\

        CoDeR-Union+BGE Reranker
        & 80.25 & \textbf{4.45}
        & 78.89 & \textbf{3.57}
        & 75.16 & \textbf{3.95}
        & 60.19 & \textbf{4.28}
        & \textbf{86.06} & \textbf{4.84} \\
        \bottomrule
    \end{tabular}
    }
    \caption{Reranking comparison across Antonym, Negation, Exclusion, NevIR, and ExcluIR using nDCG@5 and FVR.}
    \label{tab:ablation_reranker_comparison}
\end{table*}

Table~\ref{tab:ablation_reranker_comparison} tests a different alternative explanation: perhaps a standard reranker can repair constraint violations after retrieval, making compatibility-aware retrieval unnecessary. This ablation is aligned with the methodology in which CoDeR operates before optional downstream rerankers or generators. The question is therefore not whether a BGE reranker can improve semantic ranking quality, but whether a topical reranker can undo the exposure of constraint-violating evidence that has already entered the candidate list.

The comparison suggests that reranking and compatibility-aware retrieval address different parts of the pipeline. BGE-large plus a BGE reranker remains strong in nDCG because both components reward semantic usefulness and topical match. However, this does not guarantee a later first violation, especially on settings such as Negation, Exclusion, and ExcluIR where the violating evidence is intentionally topically close to the satisfying evidence. In contrast, applying the same reranker after CoDeR starts from a candidate list whose risk structure has already been changed by compatibility scoring. The reranker can then preserve or refine relevance without having to discover the constraint direction from scratch.

This result clarifies the role of CoDeR in a RAG-style pipeline. CoDeR is not proposed as a replacement for all downstream reranking; it is a retrieval-side mechanism for reducing harmful early exposure before later components consume the context. The ablation therefore supports the broader claim of the paper: constraint compatibility is not a by-product of stronger topical retrieval or standard reranking. It must be introduced as an explicit retrieval signal that separates topically plausible satisfying evidence from topically plausible violations.

\section{Preliminary RAG-Oriented Probe}
\label{app:end2end}

To connect retrieval-side violation control with answer-level behavior, we conduct a lightweight End-to-End RAG stress test on 20 randomly sampled NevIR question-style queries.
We evaluate whether the top retrieved evidence supports the satisfying answer direction, while keeping the corpus, queries, retrieval depth, generator, and prompt fixed across methods.

\begin{table}[t]
    \centering
    \small
    \begin{tabular}{l|cc}
        \toprule
        Method & Answer Accuracy Rate$\uparrow$ & FVR Avg.$\uparrow$ \\
        \midrule
        BM25 & 15.00\% & 2.90 \\
        BGE & 30.00\% & 3.45 \\
        Contriever & 25.00\% & 3.15 \\
        \specialrule{0.25pt}{2pt}{0pt}
        HyDE & 30.00\% & 3.00 \\
        NS-IR & 25.00\% & 3.45 \\
        DEO & 30.00\% & 3.35 \\
        \specialrule{0.25pt}{2pt}{0pt}
        CoDeR-Seq & \textbf{40.00\%} & \textbf{3.80} \\
        CoDeR-Union & 35.00\% & \textbf{3.80} \\
        \bottomrule
    \end{tabular}
    \caption{
    End-to-end RAG stress test on 20 randomly sampled NevIR queries.
    }
    \label{tab:end2end_nevir}
\end{table}

Table~\ref{tab:end2end_nevir} is a small downstream probe rather than a full RAG benchmark.
It connects FVR to generation risk: early violations can ground the generator in evidence that is topically relevant but directionally wrong.
CoDeR delays this context contamination and more often places satisfying evidence before conflicting evidence.
These preliminary probe results are consistent with the retrieval-side trend.

\subsection{Downstream Probe Prompt}
\label{app:e2eprompts}

\begingroup
\setlength{\fboxsep}{6pt}
\setlength{\fboxrule}{0.5pt}
\noindent
\fbox{
\begin{minipage}{0.94\linewidth}
\footnotesize
\ttfamily
\raggedright
You are a question answering assistant.\\[0.6em]

You are given retrieved passages that are relevant to the question.\\[0.6em]

Answer the question based on the retrieved passage.\\[0.8em]

Question:\\
\{query\}\\[0.8em]

Retrieved passage:\\
\{retrieved\_doc\}\\[0.8em]

Rules:\\
- Output only the short answer.\\
- Do not output any explanation, punctuation, or extra text.
\end{minipage}
}
\endgroup

\end{document}